\PassOptionsToPackage{dvipsnames}{xcolor}
\documentclass[submission,letterpaper, Phys]{SciPost}

\pdfoutput=1
\usepackage{amsmath,amssymb,mathtools,xspace}
\usepackage{booktabs,multirow,graphicx,tabularx,slashed}
\usepackage{hyperref}
\usepackage[normalem]{ulem}
\usepackage{enumitem}
\usepackage{feynmp}
\usepackage{braket}
\usepackage{tikz}
\usetikzlibrary{shapes.geometric, arrows}
\usepackage{relsize}
\usepackage{cleveref}
\usepackage{mhchem}
\usepackage[utf8]{inputenc}

\makeatletter
\@ifundefined{pdfoutput}{}{\DeclareGraphicsRule{*}{mps}{*}{}}
\makeatother

\tikzstyle{generator} = [rectangle, rounded corners, minimum width=3cm, minimum height=1cm,text centered, draw=Gcolor]
\tikzstyle{discriminator} = [rectangle, rounded corners, minimum width=3cm, minimum height=1cm,text centered, draw=Dcolor]
\tikzstyle{io} = [circle, trapezium left angle=70, trapezium right angle=110, minimum width=1cm, minimum height=1cm, text centered, draw=black]

\tikzstyle{process} = [rectangle, minimum width=1cm, minimum height=1cm, text centered, draw=black]
\tikzstyle{decision} = [rectangle, minimum width=1cm, minimum height=1cm, text centered, draw=black]

\tikzstyle{arrow} = [thick,->,>=stealth]
\usepackage{xcolor}

\DeclareMathOperator{\br}{BR}

\DeclareMathOperator{\Tr}{Tr}

\newcommand\one{\leavevmode\hbox{\small1\normalsize\kern-.33em1}}

\newcommand{\mat}{\mathcal{M}}
\newcommand{\lag}{\mathcal{L}}
\newcommand{\ord}{\mathcal{O}}

\newcommand{\gev}{\text{GeV}}
\newcommand{\tev}{\text{TeV}}

\newcommand{\iab}{\ensuremath{\text{ab}^{-1}} }

\def\slashchar#1{\setbox0=\hbox{$#1$}           
   \dimen0=\wd0                                 
   \setbox1=\hbox{/} \dimen1=\wd1               
   \ifdim\dimen0>\dimen1                        
      \rlap{\hbox to \dimen0{\hfil/\hfil}}      
      #1                                        
   \else                                        
      \rlap{\hbox to \dimen1{\hfil$#1$\hfil}}   
      /                                         
   \fi}

\newcommand{\eg}{\textsl{e.g.}\;}
\newcommand{\ie}{\textsl{i.e.}\;}

\newcommand{\lumi}{\ensuremath{\mathcal{L}}\xspace}

\begin{document}
\begin{fmffile}{feynman}

\vspace*{-1.4cm}{\mbox{}\hfill IPPP/20/21}\\[-0.5cm]

\begin{center}{\Large \textbf{
Light Dark Matter Annihilation and Scattering in LHC Detectors
}}\end{center}

\begin{center}
Martin Bauer\textsuperscript{1},
Patrick Foldenauer\textsuperscript{1},
Peter Reimitz\textsuperscript{2}, and
Tilman Plehn\textsuperscript{2}
\end{center}

\begin{center}
{\bf 1} \textit{Institute for Particle Physics Phenomenology, Durham University, United Kingdom} \\
{\bf 2} \textit{Institut f\"ur Theoretische Physik, Universit\"at Heidelberg, Germany}
\end{center}

\section*{Abstract}
{\bf We systematically study models with light scalar and pseudoscalar
  dark matter candidates and their potential signals at the
  LHC. First, we derive cosmological bounds on models with the
  Standard Model Higgs mediator and with a new weak-scale
  mediator. Next, we study two processes inspired by the indirect and
  direct detection process topologies, now happening inside the LHC
  detectors.  We find that LHC can observe very light dark matter over
  a huge mass range if it is produced in mediator decays and
  then scatters with the detector material to generate jets in the
  nuclear recoil.}

\vspace{10pt}
\noindent\rule{\textwidth}{1pt}
\tableofcontents\thispagestyle{fancy}
\noindent\rule{\textwidth}{1pt}

\newpage
\section{Introduction}
\label{sec:intro}

Discovering the nature of dark matter remains one of the most exciting
and pressing goals of fundamental physics. Unfortunately, the mass of
the dark matter agent is not known. As a starting point one can
distinguish between dark matter with large momenta compared to the
local energy density, behaving like particles, and dark matter with
small momenta compared to the local energy density which leads to a
high occupation number and can be described as a classical wave. We
know that dark matter is non-relativistic with a velocity fixed by the
escape velocity of galaxies, so dark matter momentum translates
directly into masses. Relatively light dark matter can be fermionic or
bosonic, while dark matter with masses below few 100~eV can only be
bosonic because of the Pauli exclusion principle. It is then referred
to as ultralight dark matter (ULDM) or fuzzy dark matter. Typically,
particle and wave dark matter require very different search strategies
with different reach and limitations.

Searches for particle dark matter with direct detection experiments
have limited mass reach due to recoil energy
thresholds~\cite{Knapen:2017xzo} and for indirect detection
backgrounds become very challenging below the soft X-ray spectrum
scale as evidenced by the controversy around the $3.5~$keV
line~\cite{Bernal:2017kxu}. Collider searches for dark matter are
different from both direct and indirect detection experiments in that
they are sensitive to the mediator mass, which determines the missing
energy and are largely independent of the dark matter mass as long as
the mediator can decay on-shell. If dark matter is lighter than
roughly keV and the mediator is not heavy enough to produce a large
missing energy signal ($\lesssim 100$~GeV), it could have significant
interactions with SM particles and still escape established search
strategies for particle dark matter. New experimental techniques aim
to extend direct detection probes to smaller
masses~\cite{Battaglieri:2017aum}, by utilizing electron
recoils~\cite{Essig:2012yx,Graham:2012su,Hochberg:2017wce,Derenzo:2016fse,Cavoto:2017otc,Kurinsky:2019pgb,Hochberg:2019cyy},
beta decays and nuclear absorption~\cite{Adhikari:2016bei}, atomic
excitations~\cite{Essig:2011nj}, or absorption in superconductors and
other systems with small gaps~\cite{Hochberg:2015pha,
  Hochberg:2016ajh, Hochberg:2017wce}. Some of these projections
promise sensitivity down to the eV scale.  Below this mass scale, wave
dark matter leads to the variation of fundamental constants, which can
be probed in atomic spectroscopy~\cite{Stadnik:2015kia,
  Stadnik:2015uka, Brax:2017xho, Fichet:2017bng }, laser
interference~\cite{Stadnik:2015xbn}, Eot-Wash and ULDM-fifth-force
experiments~\cite{Hees:2018fpg, Jones:2019qny}.  Other new strategies
include nuclear clocks~\cite{Banerjee:2020kww}, compact binary systems~\cite{Poddar:2019zoe,Poddar:2020qft} or gravitational wave
detectors~\cite{Grote:2019uvn,Bertoldi:2019tck,Badurina:2019hst}.  All
these measurements probe masses down to $10^{-24}$~eV, below which the
de Broglie wavelength is larger than ~100 kpc and galaxy-size
structures do not form~\cite{Hlozek:2014lca}.

There exist a few reasons to favor wave dark matter over particle dark
matter, because the macroscopic de Broglie wave length can suppress
the formation of small structures and lead to less cuspy halo profiles
as opposed to cold dark matter~\cite{Frieman:1995pm, Hu:2000ke,
  Amendola:2005ad, Marsh:2010wq, Hui:2016ltb}. For dark matter that
only interacts gravitationally, there is a narrow, preferred mass
scale of $10^{-22}$ eV which suppresses kpc-sized cusps and substructures.
If there are additional \emph{repulsive} interactions, wave dark
matter can solve these problems for a range of masses over many orders
of magnitude~\cite{Fan:2016rda, Dev:2016hxv, Lee:2017qve}. Even though this wave
dark matter can be described classically, it is clear that there has
to exist a description in terms of quantum field theory.

In this paper we provide an overview of many constraints on ULDM,
formulated in terms of particle physics models. A proper description
of these models in terms of a quantum field theory might not be
necessary to derive certain implications of wave dark matter, but
developing such models without a microscopic theory is overly
naive~\footnote{In short, quantizing fields relevant for cosmology is
  a matter of physics honor, not of necessity.}. In
Sec.~\ref{sec:models_dm} we define a set of models for scalar and
pseudoscalar dark matter and derive their respective constraints from
cosmology and low energy physics. When we add such a light dark matter
field to the Standard Model Lagrangian and study constraints, we need
to define a mediator. An obvious choice is the Standard Model Higgs,
such that the light dark matter field becomes a minimal extension of
the scalar sector of the Standard Model. However, there is always the
chance that this mediator is a new scalar particle of an extended
Higgs sector. While we do not claim that these scalar and pseudoscalar
models give a comprehensive coverage of the light dark matter model
space~\cite{Jaeckel:2012yz,Soper:2014ska,Batell:2018fqo}, they do
provide representative cases for further studies.  

\begin{figure}[b!]
\centering
\includegraphics[width=.71\textwidth]{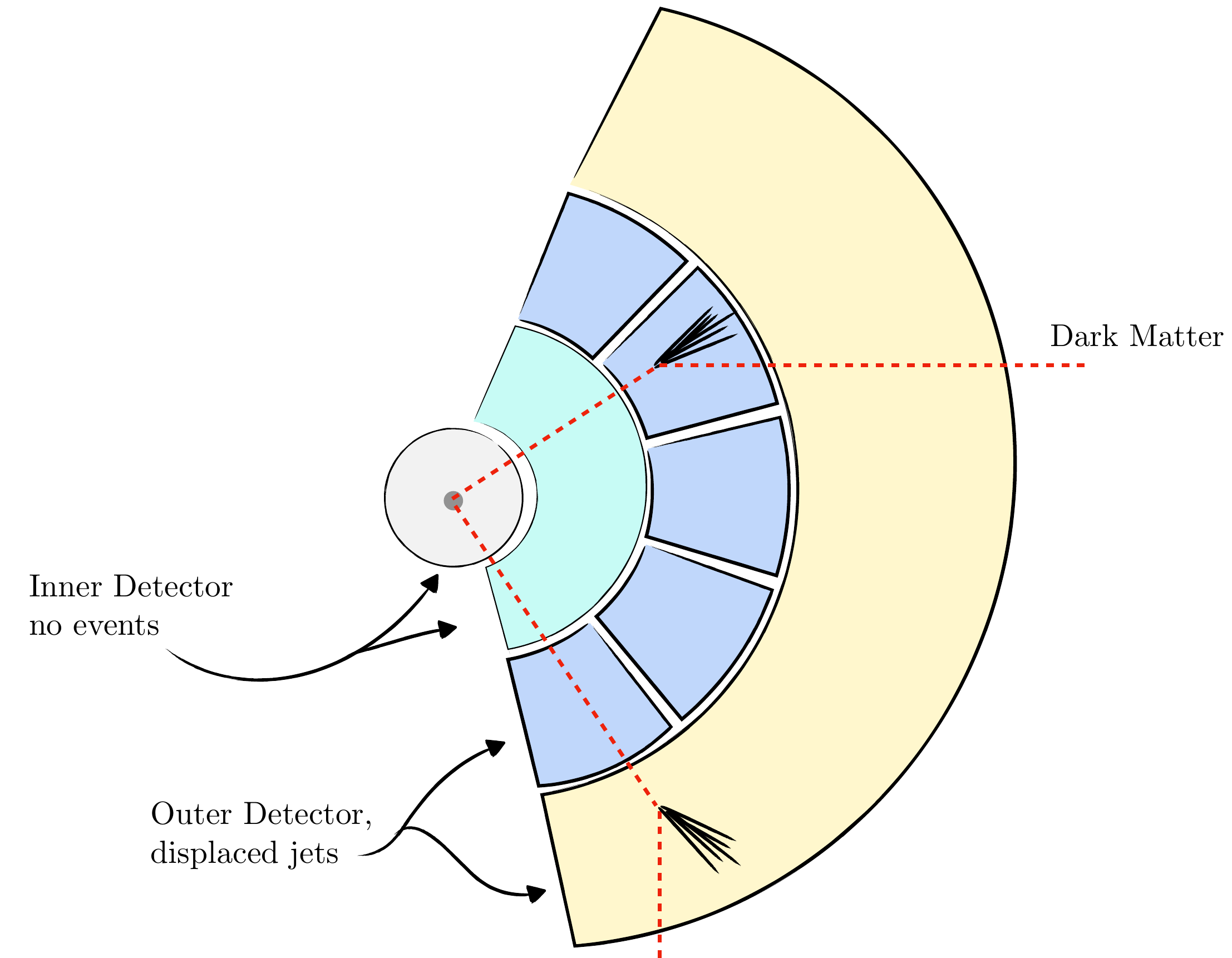}
\caption{Illustration of the displaced recoil jet signature.}
\label{fig:signature}
\end{figure}

In Sec.~\ref{sec:LHC} we then argue that collider searches can provide
competitive and complementary sensitivity reach for light and
ultra-light dark matter. We consider two processes, inspired by the
process topology of \textsl{direct and indirect detection}, but now
applied to interactions with the LHC detectors.  First, ULDM could be
produced at large boost and annihilate with the local dark matter
density in ATLAS or CMS. This would lead to the production of pairs of
photons or electrons, in analogy to indirect detection
signals. Unfortunately, we find that ATLAS cannot simply play the role
of the galactic center and provide a discovery. Second, ULDM can again be
produced for instance in Higgs decays and then inelastically scatter
off the detector material. In analogy to direct detection it will
leave displaced jets which are extremely unlikely to appear as
Standard Model backgrounds.  Such \textsl{displaced recoil jets} can
generally appear in the first dense layers of the LHC detectors, which
for ATLAS will be the two calorimeters and for CMS can include the
silicon tracker. They are fundamentally different from other displaced
signatures such as displaced vertices from decaying
massive particle or emerging jets~\cite{Schwaller:2015gea}. Unlike classic direct detection this signal does not
probe the actual dark matter nature of the candidate
particle. Nevertheless, we find that this strategy is promising and
propose a new signature of mono-jets and displaced jets in dense
detector components. This kind of signature can obviously be extended
to other models with light and very weakly interacting particles
produced at the LHC, like sterile neutrinos.

\section{Light dark matter and cosmology}
\label{sec:models_dm}

Cold dark matter with a non-relativistic velocity of $v/c\sim 10^{-3}$
and mass $m$ has a local number density of $n=(0.04\, \text{eV})^4/m$
and a de Broglie wavelength of $\lambda_\text{dB}=2\pi/(mv)$. The
occupation number then scales like
\begin{align}
  n\,\lambda_\text{dB}^3\approx 6.35 \cdot 10^5\left(\frac{\text{eV}}{m}\right)^4\,,
\end{align}
which is huge for $m \lesssim 1$~eV, so dark matter in this regime can
be treated as a classical field~\cite{Sikivie:2009qn}. In this mass
range dark matter can only be bosonic and the correct relic density
can be explained by the misalignment mechanism~\cite{Abbott:1982af,
  Preskill:1982cy} or by the Affleck-Dine mechansim if the scalar sector has an explicitly broken global symmetry~\cite{Alonso-Alvarez:2019pfe}. In this regime the only possible decay is into photons and the
decay width is suppressed by the dark matter mass. As an example, a
dark matter scalar can couple to two powers of the QED field strength
$F_{\mu\nu}$ such that its life time scales like
$m^3/\Lambda^2$. Similarly, a light vector can couple to three powers
of the field strength, leading to a life time proportional to
$m^9/\Lambda^8$. Depending on the suppression scale, this implies that
light dark matter does not require an additional symmetry to make it
stable for $\Lambda \gtrsim 10^4\,\text{GeV} (m/\text{eV})^{3/2}$ and
$\Lambda \gtrsim 2.25\,\text{keV} (m/\text{eV})^{9/8}$,
respectively~\cite{Pospelov:2008jk}. This is particularly true for the
QCD axion for which the mass is related to the suppression scale, $m
\approx 5.7\cdot 10^{-6} \,\text{eV}
(10^{12}\,\text{GeV}/\Lambda)$~\cite{diCortona:2015ldu}.  Stability
becomes a bad assumption for smaller suppression scales $\Lambda$ or
masses above the electron-positron threshold $m > 2m_e$.

For dark matter with negligible self-interactions, the balance between
gravitational attraction and quantum pressure determines the scale for
which stable structures form and for $m\approx 10^{-22}$ eV the
quantum pressure suppresses kpc structures~\cite{Hu:2000ke}. This
opens a solution to the small scale problems of cold particle dark
matter, providing a better fit to the matter power spectrum at small
scales~\cite{Du:2016zcv}. In the presence of repulsive
self-interactions, masses below the eV-scale kpc-sized structures can be
effectively suppressed~\cite{Tkachev:1986tr, Goodman:2000tg,
  Peebles:2000yy}. In contrast, attractive self-interactions have a
destabilizing effect and lead to collapsing structures such as boson
stars~\cite{Davidson:2014hfa, Guth:2014hsa}. Axions lead to
trigonometric potentials which always induce an attractive
self-interaction~\cite{Fan:2016rda}.  Nevertheless, the focus of light
dark matter models has been on axion-like candidates inspired by
string theory~\cite{Svrcek:2006yi, Arvanitaki:2009fg,Kim:2015yna}. Given
the existing constraints they require a very low explicit breaking
scale associated with a new physics sector well below the GeV scale
even for Planck suppressed couplings~\cite{Davoudiasl:2017jke}. We
discuss models with a $Z_2$-symmetry with repulsive self-interactions
and point out interesting consequences for low-energy and cosmological
constraints in a class of models with derivative couplings and
$Z_2$-symmetry.
This symmetry guarantees dark matter stability for the whole
mass range and self-interactions can be repulsive.  In our overview we
distinguish two different classes: scalar and Goldstone-boson
(axion-like) dark matter coupled through the Higgs portal as well as
through a scalar portal with a new mediator.

\paragraph{Number of effective degrees of freedom}
Before detailing the models under study in this work, we want to give some general remarks on possible constraints on light (pseudo-)scalar DM from the number of effective degrees of freedom $N_\mathrm{eff}$ in the early universe.

Depending on the strength of the interaction mediated through the Higgs or scalar portal studied in this paper, dark matter is in thermal equilibrium with the SM thermal bath in the early universe. 
If this interaction is strong enough to keep the DM in equilibrium until after the QCD phase transition has occurred,
it can contribute a significant fraction to the radiation energy density. Such a contribution is typically measured in units of the contribution of a single relativistic neutrino species $\Delta N_\text{eff}\equiv \rho(\mathrm{DM)}/\rho(\nu)$. 
The SM predicts three relativistic species in the epoch before recombination $N_\text{eff}^\text{SM}=3.046$. This is in agreement with bounds obtained by the Planck collaboration, the strongest one coming from a fit to CMB polarisation, lensing and baryon acoustic oscillation data $N_\text{eff}^\text{SM}=2.99^{+0.34}_{-0.33}$ ($95\%$ CL)~\cite{Aghanim:2018eyx}, whereas a fit to only the power spectrum yields $N_\text{eff}^\text{SM}=3.00^{+0.57}_{-0.53}$.
This assumes  the standard $\Lambda$CDM model and depends significantly on other underlying parameters, \emph{e.g.}~the helium abundance during BBN~\cite{Pisanti:2020efz}. In BSM scenarios, \emph{e.g.}~if neutrinos are allowed to decay~\cite{Chacko:2019nej, Escudero:2020ped} or if a Majoron heats the neutrino sector~\cite{Escudero:2019gvw}, the value of the fit can change considerably.
The contribution of (pseudo-)scalar dark matter is  $\Delta N_\text{eff} \approx 2$
in the case of late decoupling, $T_\text{dec}\lesssim 1$ MeV,
whereas earlier decoupling leads to contributions of $\Delta
N_\text{eff} \lesssim 0.5 $ for $T_\text{QCD}> T_\text{dec}\gtrsim 1$
MeV, and  $\Delta N_\text{eff} \lesssim 0.05 $ for  $
T_\text{dec}\gtrsim T_\text{QCD}$~\cite{Brust:2013ova,
  Chacko:2015noa}. 
It is therefore safe to assume that DM decoupling before the QCD phase transition is currently unconstrained by data.
For DM decoupling after the QCD phase transition, but at $T_\mathrm{dec} > 1$ MeV, the contribution to $\Delta N_\mathrm{eff}$ varies between 0.05 and 0.5.
Depending on the precise decoupling temperature and the specifics of the dark  sector we consider this value unfavoured but not excluded by Planck. A future measurement by the Simons Observatory~\cite{Ade:2018sbj} or CMB-S4~\cite{Abazajian:2016yjj} with their respective projected sensitivities of $\sigma(N_\mathrm{eff})=0.05$ and $\sigma(N_\mathrm{eff})=0.03$ would firmly exclude these scenarios.

The models we are studying in this work require large couplings to gluons. As a consequence the DM will be kept in 
equilibrium with the SM until after the QCD phase transition.\footnote{We want to take this opportunity to thank Simon Knapen for pointing this out to us.}
 However, once the temperature drops below  $T_\text{dec} \lesssim m_\pi$, the dark sector is in contact with the SM thermal
bath  only via ${\rm DM~DM}\leftrightarrow e^+ e^-$ or ${\rm DM~DM}\leftrightarrow \gamma
\gamma$ scattering,\footnote{In the case of a scalar mediator and
  $T_\text{dec}< T_\text{QCD}$ with couplings only to gluons, dark
  matter can also annihilate into photons through virtual pion
  intermediate states ${\rm DM~DM} \to \pi^0\pi^0 \to 4\gamma$.} which are strongly suppressed in all models we are considering.
We therefore assume that it is always possible to decouple in the window
  $m_\pi> T_\text{dec}\gtrsim 1$ MeV, where the contribution to
  $\Delta N_\mathrm{eff}$ is still not excluded by Planck.

\subsection{Scalar dark matter} 
\label{sec:models_scalar}

A scalar singlet $s$ protected by a $Z_2$-symmetry provides a UV-complete model for light dark matter 
\begin{align}
\lag \supset \frac{1}{2}\partial_\mu s \partial^\mu s-\frac{1}{2}m_s^2s^2-\frac{1}{4!}\lambda_s s^4\,.
\label{eq:singletscalar}
\end{align}
Vacuum stability requires $\lambda_s\geq 0$, which implies repulsive
self-interactions. Renormalizable couplings to the SM can be
established through the Higgs portal
\begin{align}
\lag \supset -\frac{1}{2}\lambda_{h s}\, s^2\,H^\dagger H\,.
\label{eq:higgsportal}
\end{align}
For the SM Higgs boson we remind ourselves of the effective coupling
to gluons and photons, derived from the low-energy Lagrangian
\begin{align}
\lag \supset
 \frac{g_{h\gamma\gamma}}{v}\ h F_{\mu\nu}F^{\mu\nu}
+ \frac{g_{hgg}}{v} \ h \Tr G_{\mu\nu} G^{\mu\nu} \; ,
\label{eq:higgsgluon}
\end{align}
with $g_{hgg} = \alpha_s/(12\pi)$ and $g_{h\gamma\gamma} = -
47\alpha/(72\pi)$~\cite{Carena:2012xa} in the consistent heavy top limit and integrating
out the $W$-boson at one loop.  Higgs-induced dark matter
self-interactions can be large for sizable $\lambda_{hs}$, but any
contribution can be absorbed by choosing appropriate values of
$\lambda_s$.  Scalar dark matter with a Higgs portal is effectively a
two parameter model, and both $\lambda_{hs}$ and $m_s$ need to be
independently very small to have a viable ultra-light dark matter
candidate.

\begin{figure}[t]
\includegraphics[width=.51\textwidth]{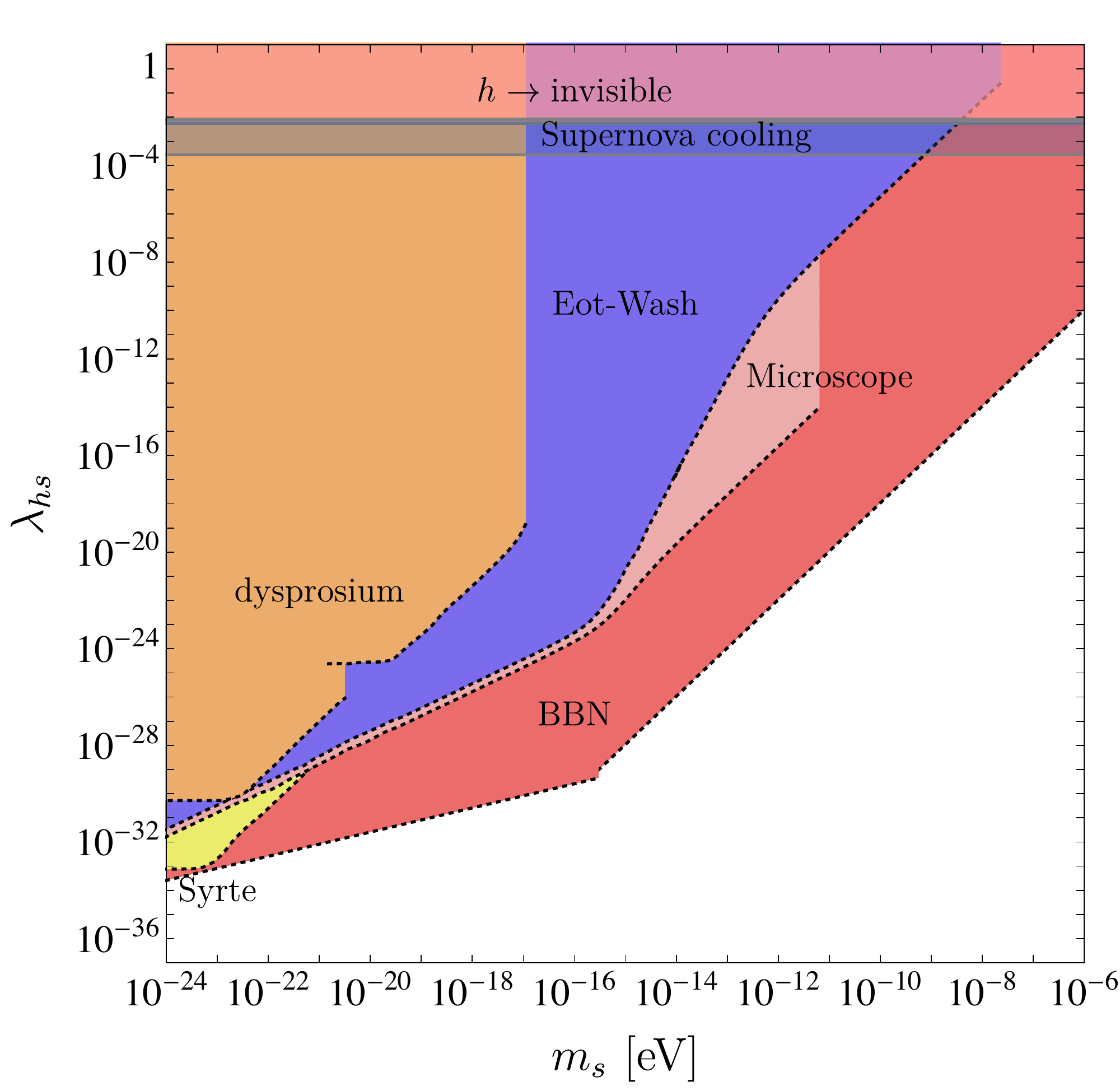}
\includegraphics[width=.5\textwidth]{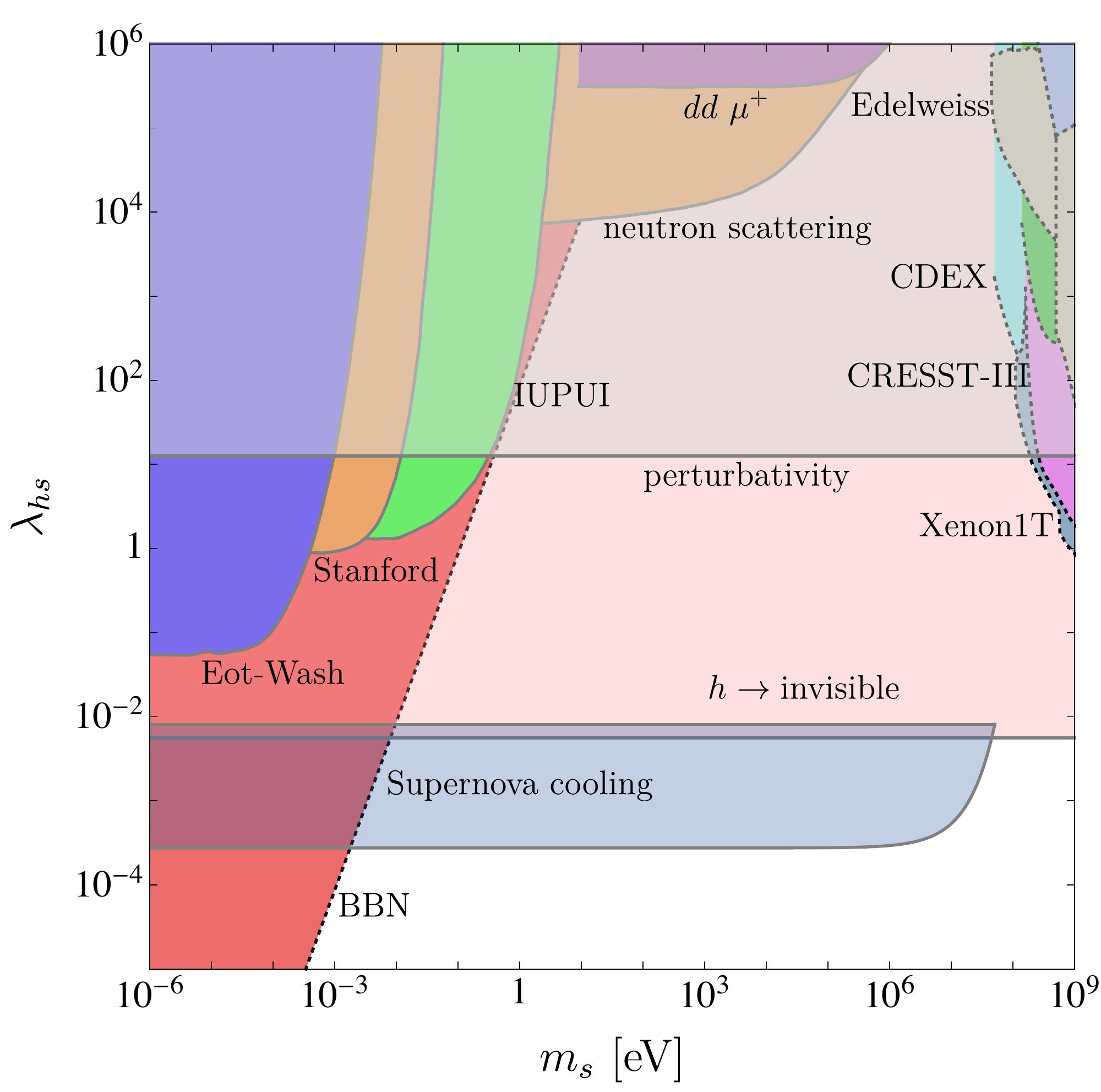}
\caption{Constraints from precision experiments, cosmology, and direct
  detection on scalar ULDM with a Higgs portal. The dark matter mass
  $m_s$ and the portal coupling $\lambda_{hs}$ are the only free
  parameters. Constraints which require the dark matter nature are
  shown with dotted contours. The perturbativity bound of
  $\lambda_{hs}=4\pi$ is shown in shaded grey.}
\label{fig:higgs-portal}
\end{figure}

\begin{figure}[t]
\includegraphics[width=.51\textwidth]{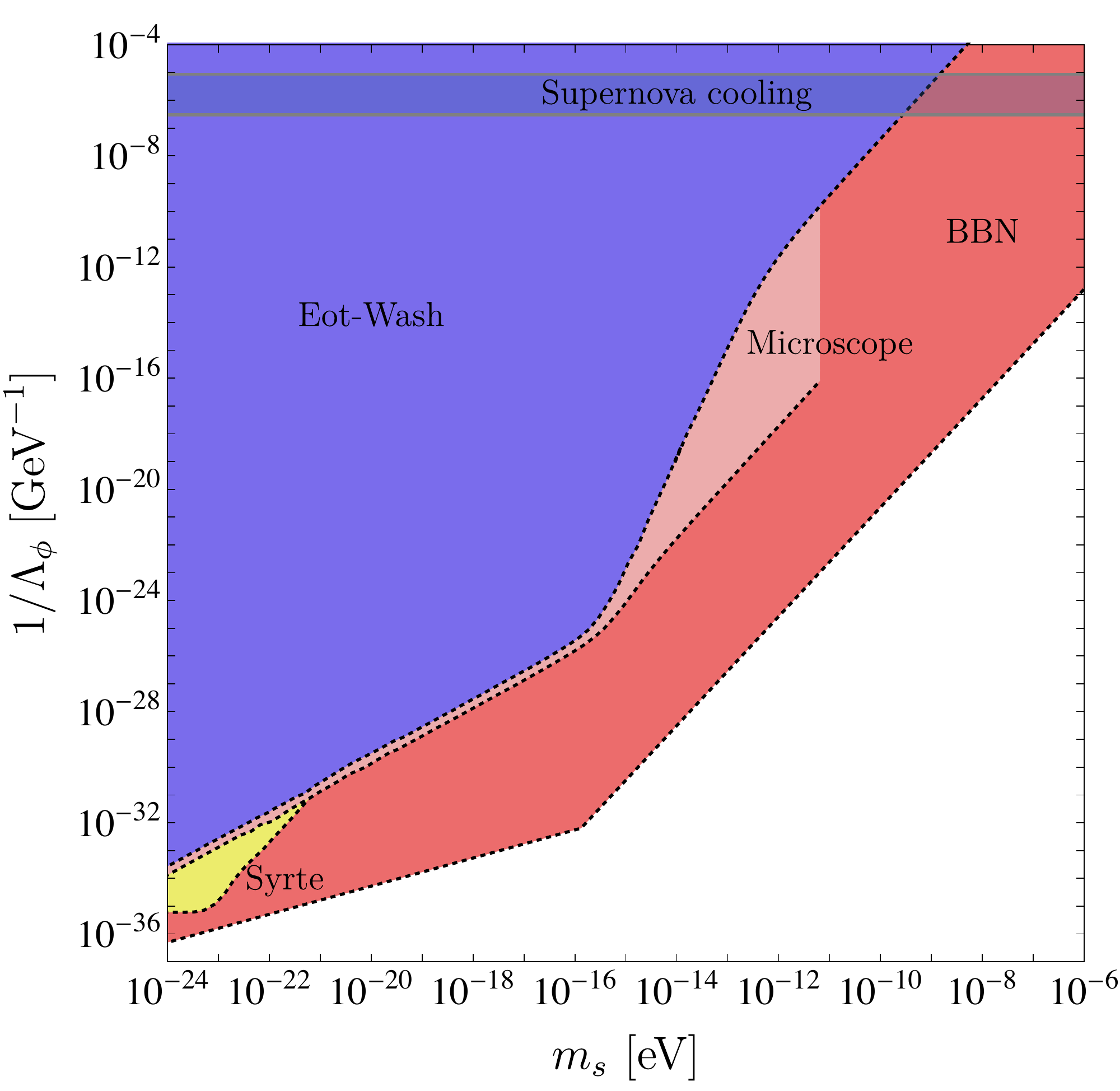}
\includegraphics[width=.5\textwidth]{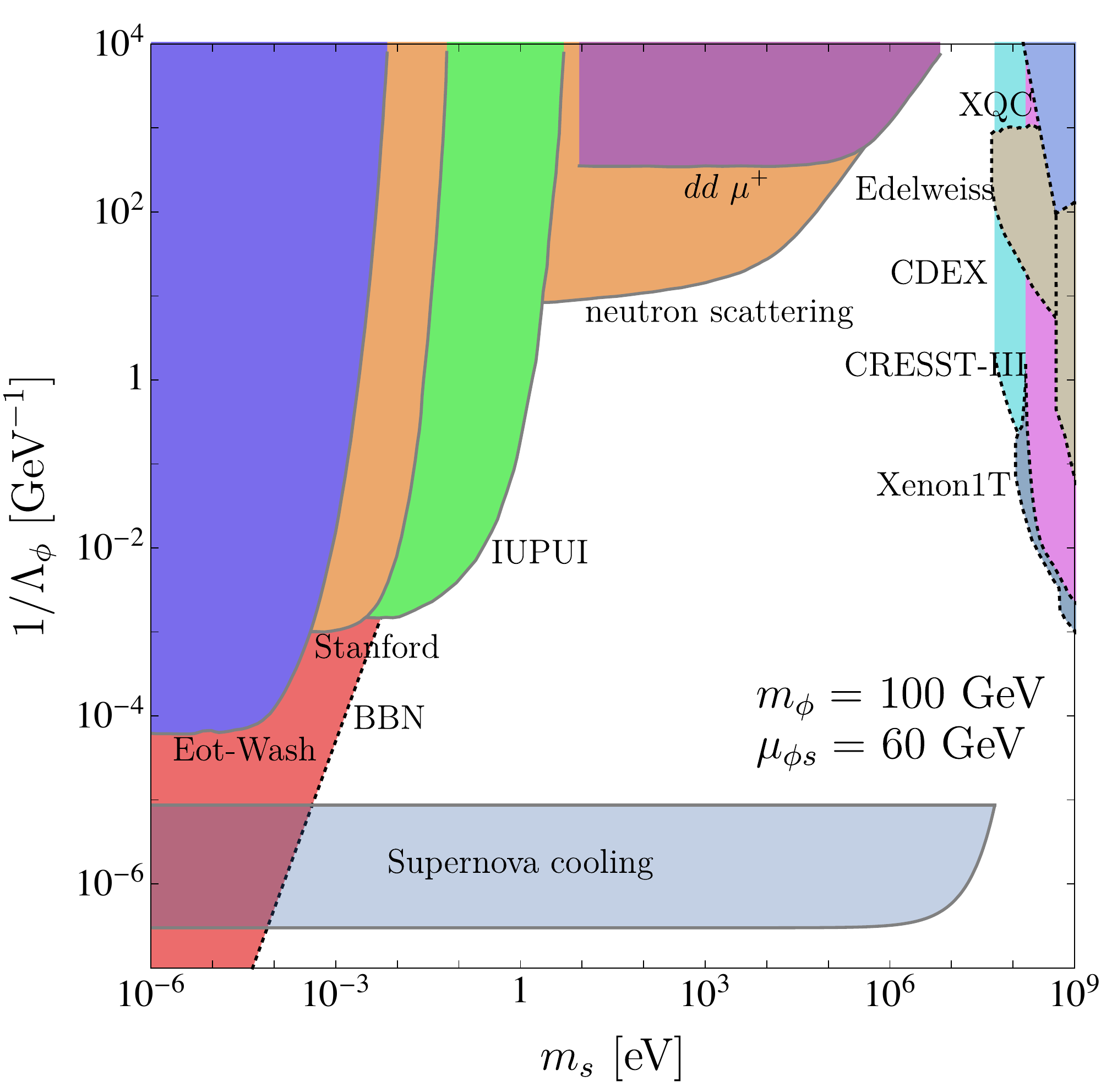}
\caption{Constraints from precision experiments, cosmology, and direct
  detection on scalar ULDM with a scalar mediator. In addition to the
  dark matter mass $m_s$ and mediator-gluon coupling $1/\Lambda_\phi$
  we fix the mediator mass to $m_\phi = 100$~GeV its coupling to the
  dark matter agent to $\mu_{\phi s} = 60$~GeV. Constraints which
  require the dark matter nature are shown with dotted contours.}
\label{fig:scalar-mediator}
\end{figure}

As an alternative, we consider the Lagrangian of
Eq.\eqref{eq:singletscalar} without a Higgs portal, but with a new
scalar mediator $\phi$ and an effective coupling to gluons
\begin{align}
\lag \supset
-\frac{1}{2}m_\phi^2\phi^2
-\frac{\mu_{\phi s}}{2}\phi s^2
-\frac{\alpha_s}{\Lambda_\phi} \; \phi \; \Tr G_{\mu\nu}G^{\mu\nu} \,.
\label{eq:scalarlag}
\end{align}
In contrast to the Higgs portal model, the mediator model introduces
three additional parameters, the mediator mass $m_\phi$, the
dimensionful coupling strength to dark matter $\mu_{\phi s}$ and a
coupling to gluons suppressed by the scale $\Lambda_\phi$.  For both
of these models we need to consider a set of low-energy and
cosmological constraints. We will collect them in
Fig.~\ref{fig:higgs-portal} for the Higgs portal and in
Fig.~\ref{fig:scalar-mediator} for the scalar mediator.  In the case of
the mediator model we usually fix $m_\phi=100$~GeV.

\paragraph{Low-energy constraints}  
The exchange of very light dark matter pairs between nuclei creates a
potential that affects precision measurements of low-energy
observables.  Constraints from neutron scattering, fifth-force
searches, Eot-Wash experiments and molecular spectroscopy
measurements can be adapted to our fundamental models from
Ref~\cite{Brax:2017xho}, where they are reported in terms of effective
couplings to nucleons. In our case, the mediator coupling to the gluon
coupling induces an effective mediator-nucleon interaction the same
way it does for the Higgs~\cite{Gunion:1989we},
\begin{align}
\lag \supset g_{\phi NN}\ \bar{N} N \,\phi \,,
\label{eq:nucl_cpl}
\end{align}
where the effective coupling in terms of the QCD partons is 
\begin{align}
g_{\phi NN} = \frac{8\pi}{11- \frac{2}{3} n_L}\, \frac{m_N}{\Lambda_\phi} \; ,
\end{align}
and $n_L$ denotes the number of light quarks. Combined with $\mu_{\phi
  s}$ this coupling induces a contact interaction of two DM scalars
$s$ with two nucleons $N$.  We can integrate out the Higgs as well as
the scalar mediator using a matching condition of the kind
\begin{align}
\frac{g_{\phi NN}\, \mu_{\phi s}}{m_\phi^2}\, \bar N N \  \frac{s^2}{2}
= \frac{1}{\Lambda} \, \bar N N \  \frac{s^2}{2}\,.
\end{align}
and formulate limits in both scalar dark matter models in terms of
\begin{align}
\lag \supset c_{sNN} s^2 \bar N N 
\label{eq:nucluoncoupling}
\end{align}
with the dimensionful coefficients 
\begin{alignat}{7} 
c_{sNN} &= 
\lambda_{h s} \frac{m_N}{m_h^2}\frac{2n_H}{3(11-\frac{2}{3}n_L)}
&\quad &\text{(Higgs portal)} \notag \\
c_{sNN} &=  \frac{\mu_{\phi s}}{\Lambda_\phi} \frac{m_N}{m_\phi^2}\frac{8\pi}{11-\frac{2}{3}n_L}
&\quad &\text{(scalar mediator)} \; .  
\label{eq:ScalarNN}
\end{alignat}
These operators lead to a new, fifth force, mediated by the light
scalar $s$ that is independent of its dark matter character.  The
low-energy limits are shown in Fig.~\ref{fig:higgs-portal} as a
function of the dark matter mass and the Higgs portal coupling.
Eot-Wash limits on fifth forces based on data from torsion balance
experiments~\cite{Schlamminger:2007ht, Adelberger:2009zz,
  Wagner:2012ui} are shown in blue. For relatively high masses, shown
in the right panel of Fig.~\ref{fig:higgs-portal}, they lose
sensitivity for masses $m_s\gtrsim10^{-4}$~eV, corresponding to around
$10^{-4}$~m, the length scale tested by the experiment.  Searches for
fifth forces with planar geometry are less sensitive, but can probe
distances down to a few $\mu$m. The limits from experiments by the
Stanford group~\cite{Smullin:2005iv} and at Indiana-Purdue
(IUPUI)~\cite{Chen:2014oda} are shown in the right panel of
Fig.~\ref{fig:higgs-portal} in orange and green, respectively.  In
constrast, the MICROSCOPE satellite~\cite{Touboul:2017grn} tests this
interaction in the presence of DM-background and probes deviations in
the orbits of test masses. These limits, shown in light red in
Fig.~\ref{fig:higgs-portal}, are strong up to $m_s\approx10^{-12}$~eV
or length scales around $2\cdot 10^5$~m, roughly corresponding to the
orbit of the satellite~\cite{Hees:2018fpg}. The same DM-background
also leads to improved Eot-Wash limits in the same mass range.  Below
the $\mu$m scale, constraints can be set by neutron-nucleon
scattering~\cite{Nesvizhevsky:2007by}. The corresponding limit is
shown in orange in the right panel of Fig.~\ref{fig:higgs-portal}.
Finally, molecular spectroscopy experiments are sensitive to forces
below the keV scale or $10^{-10}$~m~\cite{Brax:2017xho}. The strongest
limit is provided by measurements with muonic molecular deuterium
ions~\cite{Salumbides:2015qwa} shown in purple in the right panel of
Fig.~\ref{fig:higgs-portal}.

The dark
matter halo acts like a classical background field inducing
oscillating variations in fundamental constants that would not appear
for a fifth force mediator.  From a model perspective, searches for a
fifth force and variations of fundamental constants from the dark
matter halo constrain the same parameter space.  Spectroscopy searches
are sensitive to time-dependent oscillations of nucleus and electron
masses and the fine-structure constant discussed in
App.~\ref{app:fundamental}.  Since the frequency of these oscillations
are related to the mass, $\omega = m_s c^2/\hbar$, the sensitivity of
these searches peaks for a mass related to the total measurement time, and the experiment looses sensitivity below
the lowest frequency for which one full oscillation can
be measured and for frequencies higher than the shortest time between measurements~\cite{VanTilburg:2015oza}. The strongest constraint comes
from measurements with rubidium and cesium at
LNE-SYRTE~\cite{Hees:2016gop, Hees:2018fpg}, shown in yellow in the
left panel of Fig.~\ref{fig:higgs-portal}.

\paragraph{BBN constraints} 
Many of the constraints from low-energy experiments discussed above
are not specific for dark matter. For light particles forming dark
matter the misalignment mechanism leads to time-dependent
oscillations, which induces variations in fundamental constants such
as the fine-structure constant and masses of fermions and vector
bosons, which are specific to dark matter. Big-bang nucleosynthesis (BBN)
predicts the yield of $^4$He as $Y_\text{th} = 0.24709 \pm
0.00025$~\cite{Tanabashi:2018oca}, which agrees well with the measured
value $Y_\text{exp}= 0.245 \pm 0.003$~\cite{Cyburt:2015mya} and
any variation is constrained to the range
\begin{align}
\frac{\Delta Y}{Y} = -0.008458 \pm 0.012183\,.
\label{eq:variation_Y}
\end{align}
The $^4$He abundance at the time of BBN can be written in terms of the
proton-neutron ratio
\begin{align}
Y_\text{BBN}=\frac{2}{1+\frac{p_\text{BBN}}{n_\text{BBN}}} \; ,
\label{eq:YBBN}
\end{align}
where
\begin{align}
\frac{n_\text{BBN}}{p_\text{BBN}}=\frac{n_We^{-\Gamma_n t_\text{BBN}}}{p_W+n_W(1-e^{-\Gamma_n t_\text{BBN}})}
\qquad \text{and} \quad 
\frac{n_W}{p_W}=e^{-Q_{np}/T_F} \; .
\end{align}
The neutron-proton mass difference can be approximated as $Q_{np} =
m_n-m_p=m_d- m_u + \alpha \Lambda_\text{QCD}$, and the weak freeze-out
temperature is given by~\cite{Gorbunov:2011zz}
\begin{align}
T_F & = \frac{b
      m_W^{4/3}\sin^{4/3}(\theta_W)}{\alpha^{2/3}M^{1/3}_{\text{Planck}}}
      \approx 0.75\, \text{MeV} \; .
\label{eq:QandT}
\end{align}
For small values of $\Gamma_n t_{BBN}$ a possible deviation of the
Helium yield can be traced as
\begin{align}\label{eq:deltaY}
\frac{\Delta Y}{Y} \approx \frac{\Delta(n/p)_W}{(n/p)_W}-\Delta \Gamma_n t_{BBN}\,.
\end{align}
It is sensitive to variations of proton and neutron masses, variations
of $M_W$ and $M_Z$ as well as of $\alpha$. In the case of the Higgs
portal, these variations are universal apart from the loop-induced
photon coupling and following the effective operator analysis in
Ref.~\cite{Stadnik:2015kia}, we derive the constraint for the Higgs
portal
\begin{align}
  \frac{1}{m_s^2}\frac{\lambda_{\phi
     s}}{m_h^2}\left[-0.25(2g_{h\gamma\gamma})-3.79\right]\simeq
     (-2.6\pm 3.7) \cdot 10^{-20}~\text{eV}^{-4}\,,
\label{eq:scales1_bbn}
\end{align}
for $m_s\gg 10^{-16}$~eV and
\begin{align}
\frac{1}{m_s^2}\left(\frac{m_s}{3\cdot
  10^{-16}~\mathrm{eV}}\right)^{3/2}\frac{\lambda_{h
  s}}{m_h^2}\left[-0.25
  (2g_{h\gamma\gamma})-3.79\right]
\simeq (-1.3\pm 1.8) \cdot
  10^{-20}~\mathrm{eV}^{-4}\,,
\end{align}
for $m_s\ll 10^{-16}$~eV. 

For the scalar mediator the operators in Eq.\eqref{eq:scalarlag} lead to
a universal correction to the nucleon masses induced by the coupling
to gluons, whereas corrections to other fundamental constants are
strongly suppressed. The constraint can therefore be expressed as
\begin{align}
 \frac{1}{m_s^2}\frac{4\pi}{9}\frac{\mu_{\phi s}}{\Lambda_\phi}\frac{1}{m_\phi^2} \simeq (4.2\pm 6.2)\cdot 10^{-21}\,\text{eV}^{-4}\,,
\label{eq:variation}
\end{align}
for $m_s\gg 10^{-16}$~eV and
\begin{align}
 \frac{1}{m_s^2}\left(\frac{m_s}{3\cdot
  10^{-16}~\mathrm{eV}}\right)^{3/2}\frac{4\pi}{9}\frac{\mu_{\phi s}}{\Lambda_\phi}\frac{1}{m_\phi^2} \simeq (2.1\pm 3.1)\cdot 10^{-21}\,\text{eV}^{-4}\,,
\label{eq:variation16}
\end{align}
for $m_s\ll 10^{-16}$~eV.  In Fig.~\ref{fig:higgs-portal} we see that for $m_s\lesssim
10^{-3}$~eV the observed $^4$He abundance set during BBN is indeed the
most stringent constraint. The results for a scalar mediator displayed in
Fig.~\ref{fig:scalar-mediator} show a similar situation for a fixed
mediator mass $m_\phi = 100$~GeV and coupling $\mu_{\phi s} = 60$~GeV.

\paragraph{Supernova constraints}
For masses below the supernova core temperature of
$2m_s<T_\text{SN}\approx 30$ MeV, scalar dark matter pairs can be
radiated off nuclei inside the supernova core. They leave the star and
effectively provide a new source of cooling in addition to
neutrinos~\cite{Raffelt:1996wa}. This is the so called free
streaming energy loss bound assuming that the additional particles
can escape the supernova freely. The observation of SN1987A puts a
limit on the total energy-loss rate per unit mass, volume and time of
\begin{align}
\Gamma = \varepsilon_x \rho_{\text{core}} < 10^{-14} ~\text{MeV}^5~.
\label{eq:SNlimit0}
\end{align}
The dominant process for scalar dark matter production in the core is
nucleon bremsstrahlung $NN\to NN+ ss$~\cite{Olive:2007aj}. For 
the Higgs portal and the scalar mediator this translates into
\begin{alignat}{7}
\lambda_{h s} &< 2.75 \cdot10^{-4}
&\quad &\text{(Higgs portal)} \,,\notag \\
\frac{\mu_{\phi s}}{\Lambda_\phi} &< 1.8\cdot 10^{-5}\left(\frac{m_\phi}{100\, \text{GeV}}\right)^2
&\quad &\text{(scalar mediator)} \; .
\end{alignat}
For very large couplings, the mean free path~~\cite{Zhang:2014wra}
\begin{align}
\lambda=\frac{1}{n_N(r)\,\sigma_{sN\to sN}}, \quad
n_N(r)=\begin{cases}
\dfrac{\rho_\text{core}}{m_p} & \text{for}~r\leq R_\text{SN}\,,\\[5mm]
\dfrac{\rho_\text{core}}{m_p} \left( \dfrac{R_\text{SN}}{r} \right)^m & \text{for}~r>R_\text{SN}\,,\end{cases}
\end{align}
with $m=3~...~7$ and
\begin{align}
  \sigma_{sN\to sN} \simeq \frac{1}{4\pi}c_{sNN}^2\,,
  \qquad \text{assuming} \quad m_s, E_s \ll m_N \,,
\end{align}
becomes smaller than the supernova radius $\lambda \leq R_\text{SN}$.
In that case the cooling is not efficient because of repeated
scattering of dark matter in the star~\cite{Turner:1987by,
  Raffelt:1996wa}, so the light scalars start to thermalize.  Once the
assumption of free-streaming breaks down and the dark matter is
trapped in the supernova, it builds up a scalarsphere of radius
$r_0>R_\text{SN}$ characterized by the optical depth criterion
\begin{align}
\tau_s=\int_{r_0}^{\infty}\lambda^{-1}\text{d}r=\int_{r_0}^\infty
\sigma_{sN\to sN}n_N(r) \text{d}r \leq \frac{2}{3} \; ,
\end{align}
similar
to the axiosphere~\cite{Turner:1987by,Raffelt:1996wa}. Combined with
applying the energy-loss bound of SN1987A on the black-body radiation
of the scalarsphere with a temperature profile of~\cite{Zhang:2014wra}
\begin{align}
T(r)=T_\text{SN}\left(\frac{R_\text{SN}}{r}\right)^{m/3}\,,
\end{align}
we obtain the constraints
\begin{alignat}{7}
\lambda_{h s} &< 8.1 \cdot10^{-3}
&\quad &\text{(Higgs portal)}\,, \notag \\
\frac{\mu_{\phi s}}{\Lambda_\phi} &< 5.2\cdot 10^{-4}\left(\frac{m_\phi}{100\, \text{GeV}}\right)^2
&\quad &\text{(scalar mediator)} \; .
\end{alignat}
The effect of this constraints on our model parameters is shown in
Figs.~\ref{fig:higgs-portal} and~\ref{fig:scalar-mediator}.  We note
that for DM masses close to the core temperature, $m_s \lesssim
T_\text{SN}$, we have to include a Boltzmann factor $\exp(-2\,
m_s/T_\text{SN})$ into the energy loss rate to model the temperature
dependence~\cite{Knapen:2017xzo}.  For masses above $m_s > 10^{-3}$~eV
the BBN constraints vanish and supernova cooling yields the leading
constraints, independent of the couplings and for dark matter masses
up to $m_s \sim 50$~MeV.

\paragraph{Invisible Higgs decays} 
A largely model-independent bound comes from searches for invisible Higgs
decays.  The strongest current limit on the branching ratio of Higgs
bosons into invisible final states is $\br(h \to \text{inv})\lesssim
13\%$~\cite{ATLAS:2020cjb}.  Minimizing the underlying assumptions and
interpreting the LHC limits in an effective theory framework hardly
changes this limit~\cite{Biekotter:2018rhp}.  A future high-luminosity
run of the LHC (HL-LHC) could improve this limit by an order of magnitude
$\br(h \to \text{inv})\lesssim 2\%$~\cite{Bernaciak:2014pna}. For the
Higgs portal the partial decay rate of the Higgs to two scalars is
given by
\begin{align}
\Gamma(h \to s s)=\frac{\lambda^2_{hs}}{8 \pi}
  \frac{v^2}{m_h}\sqrt{1-\frac{4m_s^2}{m_h^2}} \,.
\label{eq:GammaHss}
\end{align}
Together with the current bound on invisible Higgs decays, this
provides an essentially $m_s$-independent  limit on the portal
coupling of light DM (i.e.~$m_s \ll m_h/2$) of
\begin{alignat}{7}
\lambda_{hs}&< 5.6\cdot10^{-3} 
&\quad &\text{(current)}\,, \notag \\
\lambda_{hs}&< 2.1\cdot10^{-3}
&\quad &\text{(HL-LHC)} \; .
\label{eq:invHiggsscalar}
\end{alignat}
This constraint is absent in the case of the scalar mediator portal.
In Fig.~\ref{fig:higgs-portal} we see this limit right above the
supernova limit.

\paragraph{Direct detection} 
Finally, direct detection experiments based on heavy noble gases have
a recoil threshold of $\sim1$ keV, which translates into a sensitivity to
dark matter masses of $m \gtrsim 1...10$ GeV.  Cryogenic calorimeter
experiments can lower the nuclear threshold to $\sim 100$ eV,
providing sensitivity down to dark matter masses of $m\gtrsim 100$
MeV.
A similar threshold has been
obtained by the space based X-ray Quantum Calorimetry Experiment (XQC)
which is sensitive to strongly interacting dark matter in this mass
range~\cite{Erickcek:2007jv}.  We use the results
from~\cite{Davis:2017noy, Fichet:2017bng} to show the constraints on
the parameters of the Higgs portal in Fig.~\ref{fig:higgs-portal}.
For DM masses of $m_s\gtrsim 100$~MeV we see the leading constraints
from Xenon1T~\cite{Aprile:2019jmx} (dark grey),
CRESST-III~\cite{Abdelhameed:2019hmk} (pink), CDEX~\cite{Liu:2019kzq}
(cyan), Edelweiss~\cite{Armengaud:2019kfj} (pale brown) and
XQC~\cite{Erickcek:2007jv} (pale blue). In
Fig.~\ref{fig:scalar-mediator} we show the corresponding limits for
a scalar mediator.

{It is worth pointing out that further bounds exist from cosmic ray propagation~\cite{Cyburt:2002uw,Cappiello:2018hsu} and cosmology~\cite{Ali-Haimoud:2015pwa,Gluscevic:2017ywp,Xu:2018efh,Slatyer:2018aqg}, which constrain very large DM-nucleus cross sections.}

\subsection{Pseudoscalar dark matter}
\label{sec:models_axion}

Unlike scalar dark matter, pseudoscalar or axion-like (ALP) dark
matter~\cite{Jaeckel:2010ni} is described by a non-renormalizable Lagrangian
\begin{align}
\lag \supset 
& \frac{1}{2}\partial_\mu a \partial^\mu a-\frac{m_a^2}{2}a^2
 +\frac{\partial_\mu a}{f}\sum_i \frac{c_{i}}{2}\,\bar \psi_i \gamma_\mu\gamma_5 \psi_i \notag \\
&+c_G \frac{g_s^2}{16\pi^2}\frac{a}{f}\, \mathrm{Tr}[G_{\mu\nu}\widetilde G^{\mu\nu}]
 +c_W \frac{g^2}{16\pi^2}\frac{a}{f}\,\mathrm{Tr}[W_{\mu\nu}\widetilde W^{\mu\nu}]
 +c_B \frac{g^{\prime\,2}}{16\pi^2}\frac{a}{f}\,B_{\mu\nu}\widetilde B^{\mu\nu}\; .
\label{eq:lagaxi}
\end{align}
All pseudoscalar couplings are suppressed by at least  one power of the
mass scale $f$. To understand the role of $f$ and 
compute the couplings to the Higgs sector we consider the UV-complete
theory with a complex scalar breaking a global symmetry
\begin{align}\label{eq:complexS}
S = \frac{s+f}{\sqrt{2}} \; e^{i a/f} \; .
\end{align}
In this section the scalar mode $s$ is heavy. Its mass is set by $f$,
while the mass of the pseudoscalar $a$ is proportional to some
explicit breaking of the shift symmetry parameterized by $\mu$, such
that $m_a=\mu^2/f$. A conserved $Z_2$-symmetry $S\to -S$ forbids all
dimension-5 operators. Dimension-6 operators are introduced by the
renormalizable Higgs portal and the kinetic term of the full theory
\begin{align}
\lag &\supset  \partial_\mu S \partial^\mu S^\dagger +\mu_s^2S^\dagger S-\lambda_s( S^\dagger S)^2- \frac{1}{2}\lambda_{hs}\, S^\dagger S\, H^\dagger H \; .
\end{align}
They give a scalar mass $m_s=\sqrt{2\lambda_s}f$ and lead to an
effective, derivative Higgs portal suppressed by
$1/f^2$~\cite{Weinberg:2013kea},
\begin{align}
  \lag \supset - \frac{\lambda_{hs}}{2\,m_s^2} \; \partial_\mu a \partial^\mu a \; H^\dagger H \; .
\label{eq:derHiggsportal1}
\end{align}
The derivative Higgs-portal can also be induced by a coupling between
the complex scalar and the SM through the effective operator
\begin{align}
\lag \supset \frac{(\partial_\mu S)(\partial^\mu S)^\dagger}{\Lambda_{ha}^2}H^\dagger H  =\frac{\partial_\mu a \,\partial^\mu a}{2\Lambda_{ha}^2}H^\dagger H = \frac{\partial_\mu a \,\partial^\mu a}{4\Lambda_{ha}^2} (v^2 + 2 v\,h+h^2)\; ,
\label{eq:derivativePortal}
\end{align}
where we introduce a specific suppression $1/\Lambda_{ha}$ and, in the
last step, insert the Higgs field. In principle, there can be a hierarchy of scales $f \gg \Lambda_{ha}$ and we parametrize effects through the derivative Higgs portal by $\Lambda_{ha}$ from now on. We also note that this operator will 
be generated from the Higgs portal. Alternatively, we can write it as
\begin{align}
\frac{\partial_\mu a \,\partial^\mu a}{2\Lambda_{ha}^2}H^\dagger H 
=-\frac{m_a^2a^2}{4\Lambda_{ha}^2} (v+h)^2
  -\frac{a\partial_\mu a}{2\Lambda_{ha}^2}(v\,\partial^\mu h  + h\partial^\mu h)\; ,
\label{eq:newALPs}
\end{align}
where the second term gives rise to a momentum-dependent scalar
coupling to ALP pairs.

As a simple generalization of the Higgs mediator model we again
consider a model with a new scalar mediator $\phi$. It is defined by
the operators
\begin{align}
\lag \supset
- \frac{1}{2}m_\phi^2\phi^2
- \frac{\partial_\mu a \,\partial^\mu a}{2\Lambda_{\phi a}}\phi
- \frac{\alpha_s}{\Lambda_\phi} \; \phi \; \Tr G_{\mu\nu}G^{\mu\nu} \,. 
\label{eq:scalarALPs}
\end{align}
The coupling to gluons is described by the same parameter
$\Lambda_\phi$ as in the scalar case of
Eq.\eqref{eq:scalarlag}. However, unlike in the scalar case this new
scale is supplemented with the new physics scale of the pseudoscalar
$\Lambda_{\phi a}$. Such a scalar portal operator in a $Z_2$-protected
symmetry is not very exotic. It has for example been considered to
generate a fractional contribution to the effective number of degrees
of freedom~\cite{Weinberg:2013kea}. More recently, the derivative Higgs portal has been considered in the context of missing energy signals at the LHC in~\cite{Ruhdorfer:2019utl}.

\begin{figure}[t]
\includegraphics[width=.5\textwidth]{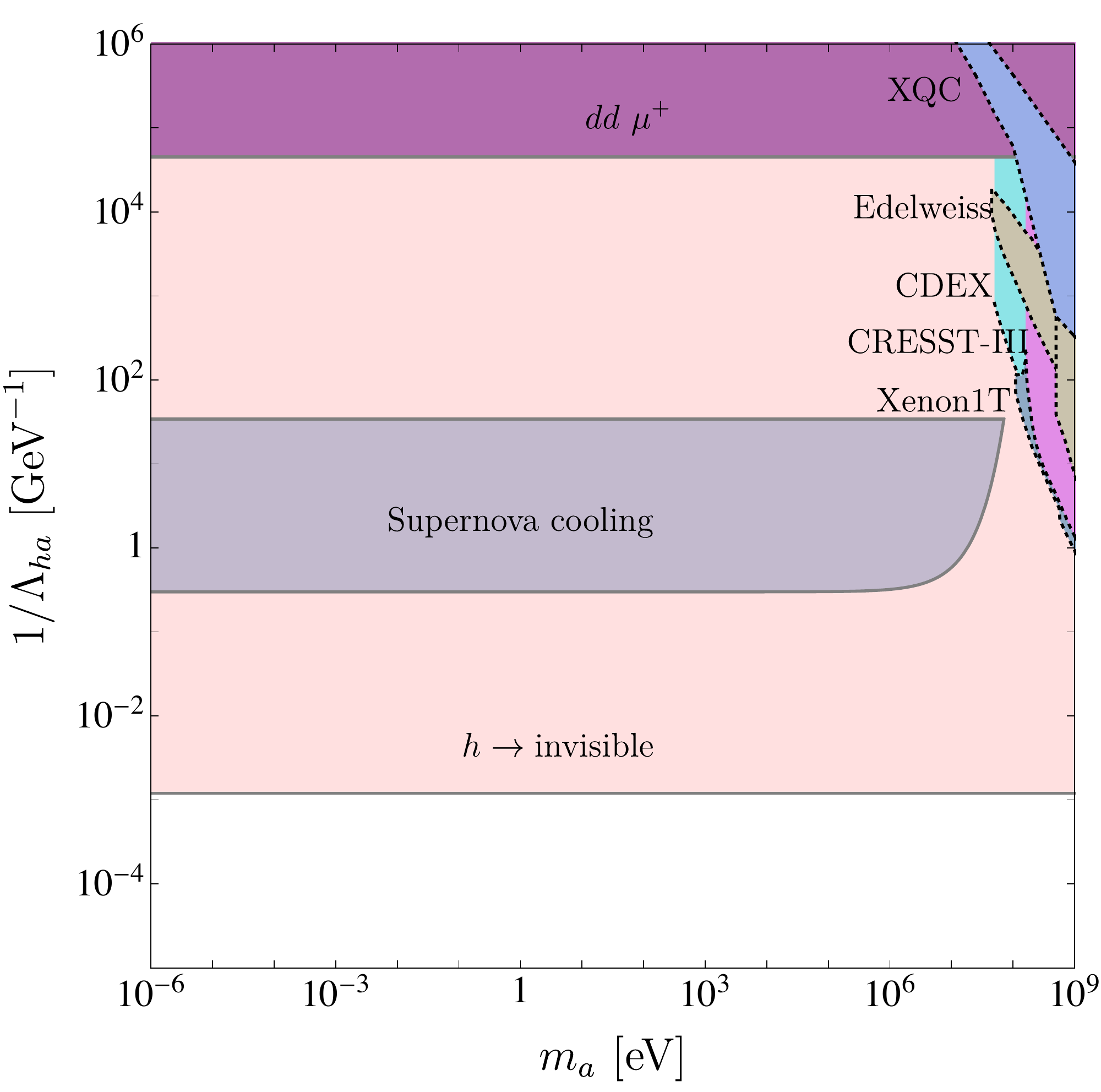}
\includegraphics[width=.5\textwidth]{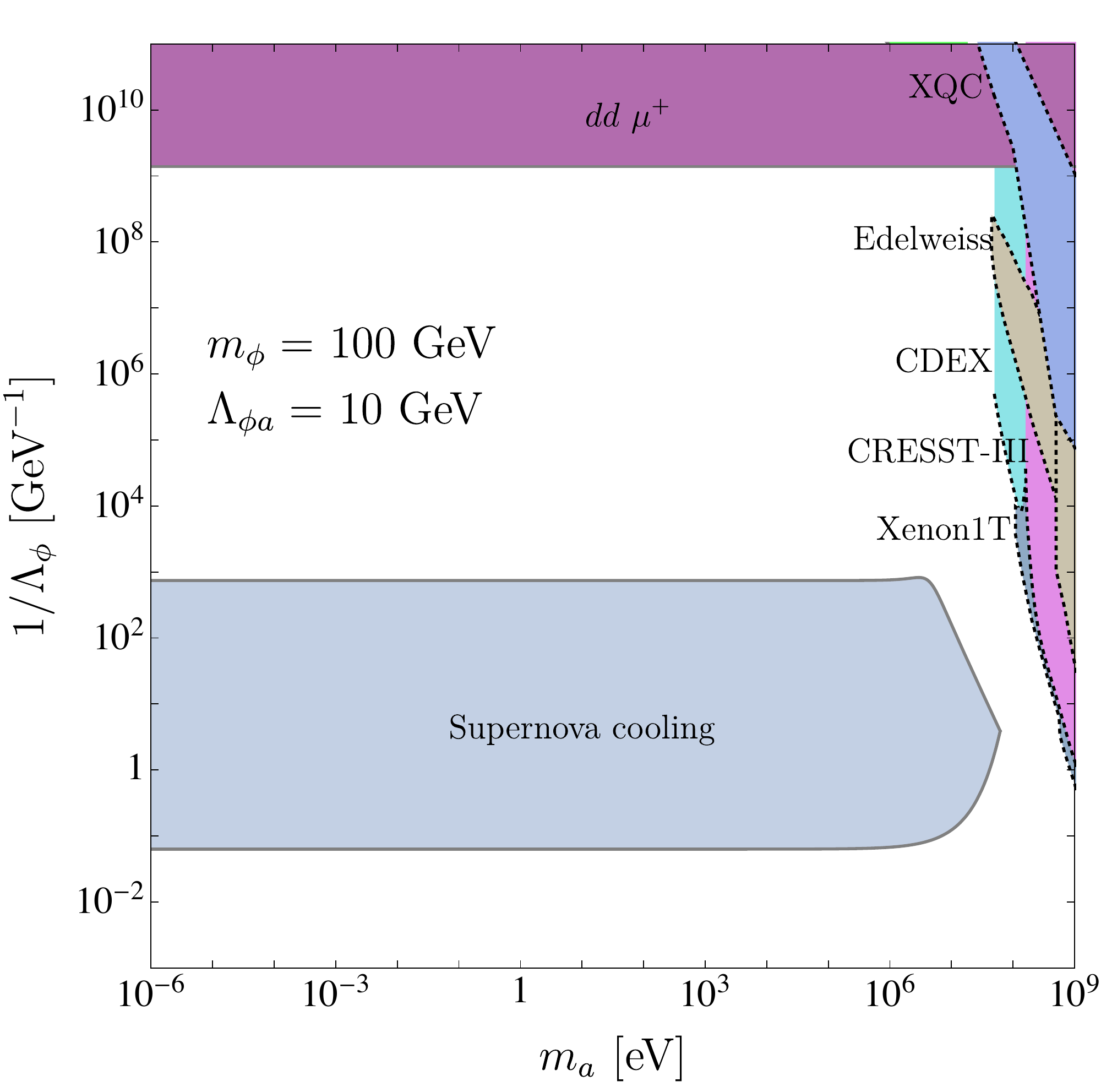}
\caption{Constraints from precision experiments, cosmology, and direct
  detector on pseudoscalar or axion-like ULDM with a Higgs mediator
  (left) and a scalar mediator (right). For the latter we again fix
  the mediator mass to $m_\phi = 100$~GeV its coupling to the dark
  matter agent to $\Lambda_{\phi a} = 10$~GeV. Constraints which
  require the dark matter nature are shown with dotted contours.}
\label{fig:alp_odd}
\end{figure}

The operators given in Eq.\eqref{eq:derivativePortal}
and~\eqref{eq:scalarALPs} induce couplings between ALPs and nuclei,
\begin{align}\label{eq:ALPNNoperator}
\lag \supset c_{aNN}\ \partial_\mu a\, \partial^\mu a\ \bar N N\,.
\end{align}
with the dimensionful coefficients
\begin{alignat}{5}
c_{aNN} &= \frac{1}{\Lambda_{ha}^2}\frac{m_N}{m_h^2}\frac{2n_H}{3(11-\frac{2}{3}n_L)}
&\quad &\text{(Higgs mediator)} \notag \\
c_{aNN} &=\frac{m_N}{\Lambda_{\phi a}\Lambda_\phi m_\phi^2} \frac{8\pi}{11-\frac{2}{3}n_L}
&\quad &\text{(scalar mediator)} \; .
\label{eq:DerNN}
\end{alignat}
Constraints from low-energy precision experiments can therefore be
discussed in analogy with the bounds on the couplings in
Eq.\eqref{eq:ScalarNN}. In Fig.~\ref{fig:alp_odd} we show the
constraints adapted from Ref.~\cite{Brax:2017xho} both for the Higgs
mediator and a new scalar mediator. In contrast to the case of the
operators in Eq.\eqref{eq:lagaxi}, all ALP interactions mediated by
Eq.\eqref{eq:derHiggsportal1} are momentum suppressed. This
generalizes to theories with more than one derivative. The sensitivity
of low-energy observables is therefore strongly suppressed with
respect to the case of an ALP with linear interactions \emph{and} in
contrast to the scalar without a shift symmetry discussed in
Sec.~\ref{sec:models_scalar}. For a more detailed discussion of this we refer to Appendix~\ref{app:models_multi}.
 The potential for the long-range force
induced by the exchange of at least two ALPs with shift-symmetry,
Eq.\eqref{eq:ALPNNoperator}, is suppressed by $1/r^7$ \cite{Brax:2017xho}, which
suppresses the sensitivity from experiments sensitive to effects at
large scales. The bounds from Eot-wash experiments and MICROSCOPE are
not relevant for the parameter space shown in Fig.~\ref{fig:alp_odd}
and constraints from neutron scattering and molecular spectroscopy
provide the dominant low-energy constraints.

\paragraph{BBN constraints} 
Constraints on pseudoscalar dark matter from big bang nucleosynthesis
are further suppressed by the derivative coupling. Treating DM as a
classical field with $a(t)=a_0\cos(m_at)$ gives an additional factor
of $m_a^2$ that is canceling the mass dependence in
Eq.\eqref{eq:variation}. The constraints now read
\begin{alignat}{7}
  \frac{1}{2\Lambda_{ha}^2m_h^2}\left( \frac{0.25c_{h\gamma\gamma}}{4\pi}+3.79\right) &\simeq
  (2.6\pm 3.7) \cdot 10^{-20}~\text{eV}^{-4}
&\quad &\text{(Higgs mediator)} \notag \\
  \frac{4\pi}{9\Lambda_{\phi}\Lambda_{\phi a} m_\phi^2} &\simeq (4.2\pm 6.2) \cdot 10^{-19}~\text{eV}^{-4}
&\quad &\text{(scalar mediator)} \; 
\label{eq:scales1}
\end{alignat}
for $m_s\gg 10^{-16}$~eV.  We see in Fig.~\ref{fig:alp_odd} that the
derivative interaction weakens the BBN bounds to the point that they
do not even appear in the plot.

\paragraph{Supernova constraints}
Similarly, constraints from supernova cooling are strongly suppressed,
because the derivatives induce additional temperature suppression in
the nuclear bremsstrahlung rate. Using the results from
Appendix~\ref{app:SN}, we find that couplings in the range
\begin{alignat}{7}
\frac{34.1}{\text{GeV}} > &\frac{1}{\Lambda_{ha}}> \frac{0.3}{\text{GeV}}
&\qquad &\text{(Higgs mediator)}\,, \notag \\
\frac{740}{\text{GeV}} > &\frac{1}{\Lambda_{\phi}} \frac{10\,\text{GeV}}{\Lambda_{\phi a}} \; \left(\frac{100\,\text{GeV}}{m_{\phi}}\right)^2 > \frac{0.063}{\text{GeV}}
&\qquad &\text{(scalar mediator)}\, ,
\end{alignat}
are excluded by supernova cooling constraints. 

\paragraph{Invisible Higgs decays}
These relatively model-independent constraints work the same way as
for the scalar case.  The new Higgs decay comes with the partial width
\begin{align}
\Gamma(h \to a a)=\frac{v^2m_h^3}{128\pi \Lambda_{h a}^4}\left(1-\frac{2m_a^2}{m_h^2}\right)^2 \sqrt{1-\frac{4m_a^2}{m_h^2}}\,\approx\, \frac{v^2m_h^3}{128\pi\Lambda_{h a}^4}\,.
\end{align}
The limits on invisible Higgs decays translate
into 
\begin{align}
  \Lambda_{h a} \gtrsim 832~\text{GeV}\,, 
\end{align}
for the current bound and $ \Lambda_{h a} \gtrsim 1.37~\tev$ for the projected bound from the HL-LHC. This reach is clearly limited and an observation would
not yield any information on the DM character of a new light
particle. Nevertheless, in the left panel of Fig.~\ref{fig:alp_odd} we
see that the invisible Higgs decays give the leading constraint on
the model.

\paragraph{Direct detection} 
Finally, we again contrast the pseudoscalar model predictions with the
limits set by different direct detection experiments. For the Higgs
and scalar mediators we see in Fig.~\ref{fig:alp_odd} that the
different experiments systematically probe their respective model
parameter space for dark matter masses exceeding $m_a \sim
50$~MeV. Because of the momentum dependence of
\eqref{eq:ALPNNoperator}, scattering from the nuclei is suppressed by
the dark matter velocity and bounds from direct detection are
considerably weaker compared to scalar dark matter. The limits from
CRESST~\cite{Angloher:2017sxg},
XQC~\cite{Erickcek:2007jv}, Xenon1T~\cite{Aprile:2019jmx},
CRESST-III~\cite{Abdelhameed:2019hmk}, CDEX~\cite{Liu:2019kzq}
Edelweiss~\cite{Armengaud:2019kfj} and XQC~\cite{Erickcek:2007jv} are
shown with the same color coding as in
Fig.~\ref{fig:scalar-mediator}. The constraints seem more important in
comparison to the supernova bounds, because the temperature
suppression in the latter is more effective than the velocity
suppression in ALP-nucleus scattering.

\section{LHC signatures of light dark matter}
\label{sec:LHC}

To look for a light new particle with a $Z_2$-symmetry and a coupling
to the Higgs sector at the LHC, we usually rely on invisible Higgs
decays. From Sec.~\ref{sec:models_dm} we know that this
process has an impressive discovery potential with very few underlying
assumptions.  However, an observation
of invisible Higgs decays would not link the new particle to dark
matter. To show such a link we could for instance search for invisible
Higgs decays where one of the two light scalar interacts with the DM
background and produces two Standard Model states. This process is the
LHC equivalent to indirect detection and will be covered in
Sec.~\ref{sec:LHC_IDD}. Alternatively, a light scalar produced in
Higgs decays can scatter with the detector, a DIS-like process which
corresponds to direct detection. We will look at it in
Sec~\ref{sec:LHC_DD}.

Before we study the potential LHC signatures we remind ourselves of
the different models defined in Sec.~\ref{sec:models_dm}:
\begin{itemize}
  \item[--] scalar $s$ with a Higgs portal, Eq.\eqref{eq:higgsportal},
    described by the renormalizable coupling $\lambda_{hs}$;
  \item[--] scalar $s$ with a mediator $\phi$,
    Eq.\eqref{eq:scalarlag}. The mediator couples to the dark matter
    scalar through $\mu_{\phi s}$ and to gluons at dimension five,
    defining $1/\Lambda_\phi$;
  \item[--] pseudoscalar $a$ with a dimension-6 coupling to the
    Higgs given by $1/\Lambda_{ha}^2$, Eq.\eqref{eq:derivativePortal};
  \item[--] pseudoscalar $a$ with a mediator $\phi$,
    Eq.\eqref{eq:scalarALPs}. The mediator coupling to the pseudoscalar is
    $1/\Lambda_{\phi a}^2$ and its coupling to gluons defines
    $1/\Lambda_\phi$;
\end{itemize}
As discussed before, these four models are subject to a wealth of
cosmological constraints. 
    
\subsection{Dark matter annihilation}
\label{sec:LHC_IDD}

\begin{figure}[b!]
\begin{center}
\includegraphics[width=0.5\textwidth]{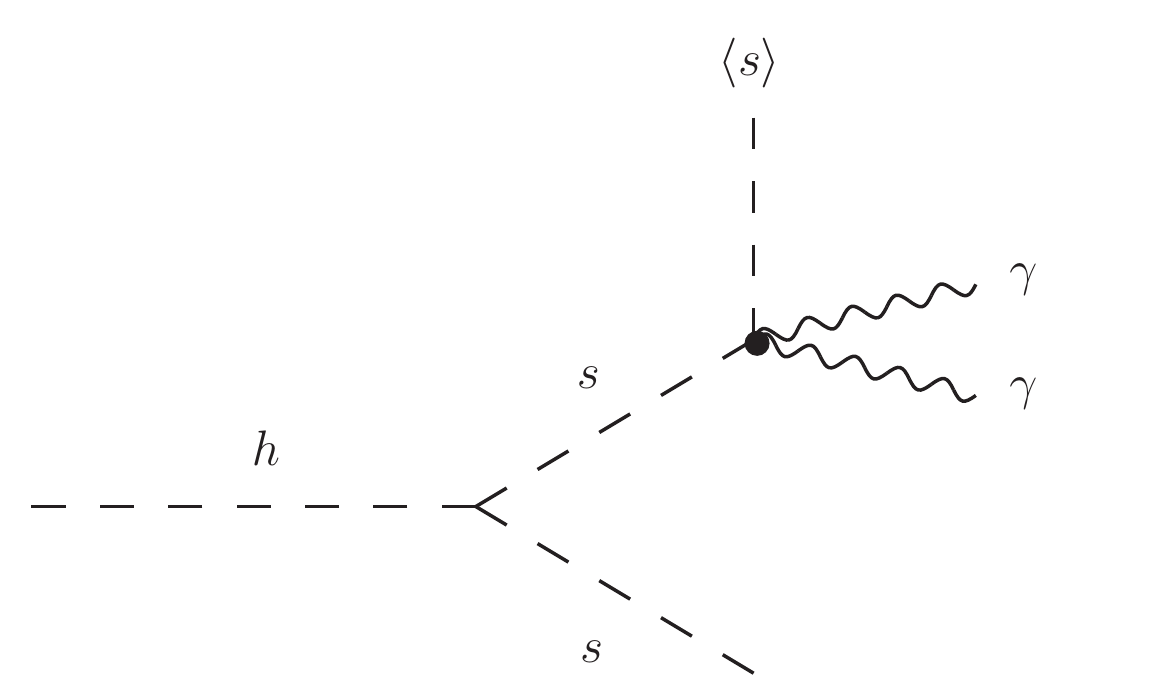}
\end{center}
\vspace{-.5cm}
\caption{Appearing pair of boosted photons
  from DM-background scattering. The produced DM scalars originate
  from a Higgs decay and share the Higgs rest mass between them.}
\label{fig:bgr_scatterAA}
\end{figure}

If the observed relic density is given by light scalars with a
$Z_2$-symmetry, a particle which we produce at the LHC can annihilate
with the dark matter background
\begin{align}
\langle s \rangle s \to \gamma \gamma \; ,
\end{align}
where $\langle s \rangle$ denotes the dark matter background, also
shown in Fig.~\ref{fig:bgr_scatterAA}. We assume that pairs of scalars
$s$ are produced in Higgs decays and then traverse the dense DM
background at high momentum. As an example we stick to the Higgs
portal model defined by Eq.\eqref{eq:higgsportal} throughout this
section.  In analogy to fixed target experiments the number of photon
pairs produced from a beam of $N_s$ scalars $s$ with initial energy
$E_s$ at a distance $l$ from the production point can be estimated
as~\cite{Andreas:2013xxa,Marsicano:2018krp,Bauer:2018onh}
\begin{align}
\frac{d^2 N_{\gamma\gamma}}{d E_s \, dl} &= N_s\, I_s(E_0, E_s, l) \  \frac{d P_\text{conv}}{dl}\,.
\label{eq:id_diff_conv}
\end{align}
Here $I_s$ denotes the energy distribution function of the scalars
$s$. The number of produced scalars can be calculated as $N_s = N_h
\ \text{BR}(h\to ss)$.  For simplicity we assume $I_s(E_s, E_0,l) =
\delta(E_s - E_0)$, which is justified as long as we do not have
significant additional interactions of the scalars with the DM medium.
One example for such an interaction would be scalar self-interactions, but they are expected
to be small for a DM candidate because of structure formation.

Finally, we are interested in the number of DM-to-photon conversions
taking place in the fiducial detector volume.  As the scalars move
through the DM gas the conversion is described by the usual
probability of a single scalar to annihilate with a DM background
particle into two photons in a spatial slice $dl$.  The differential
conversion probability for a single incoming scalar $s$ is given by
\begin{align}
\frac{d P_\text{conv}(l)}{dl} 
= \frac{e^{-l/\lambda}}{\lambda} \; ,
\end{align}
with the mean free path
\begin{align}
\lambda = \frac{1}{n_\text{DM} \; \sigma_{\langle s \rangle s \to \gamma\gamma}} \,,
\label{eq:id_path}
\end{align}
in terms of the DM number density $n_\text{DM}$.  Integrating
Eq.\eqref{eq:id_diff_conv} along a detector with size $L_\text{det}$
at a distance $L_0$ away from the interaction point gives us
\begin{align}
N_{\gamma\gamma}   &= N_s \int_{0}^{m_h/2} d E_s \ \delta(E_s - m_h/2)\,  \int_{L_0}^{L_0+L_\text{det}}  dl' \frac{e^{-l/\lambda}}{\lambda}  \notag \\
 &=N_h \ \text{BR}(h\to ss) \, \ e^{-L_0/\lambda} \left( 1 - e^{-L_\text{det}/\lambda} \right) , \;
\end{align}
where $\lambda$ is evaluated at the Higgs mass scale and the effective
Higgs couplings are defined in Eq.\eqref{eq:higgsgluon}.  The
corresponding differential cross section is
\begin{align}
\frac{d \sigma_{\langle s \rangle s \to \gamma \gamma}}{d t} 
= \frac{1}{2\pi } \,\frac{\lambda_{hs}^2 \, g_{h\gamma\gamma}^2}{(s-m_h^2)^2} \; .
\end{align}
As for all fixed target experiments, the center-of-mass energy is much
lower than the momentum of the incoming DM particle hitting the DM
target, in our case $s =m_s m_h$. With the integration bounds of
Ref.~\cite{Tanabashi:2018oca} this leads to the typical scaling of the
total rate with the DM mass
\begin{align}
\sigma_{\langle s \rangle s \to \gamma \gamma}
\approx \frac{ \lambda_{hs}^2g_{h\gamma\gamma}^2}{4 \pi}  \, \frac{m_s}{m_h^3} \; .
\label{eq:gg_xsec}
\end{align}
We also need the local DM particle number density
$n_\text{DM}$
\begin{align}
n_\text{DM} 
= \frac{\rho_\text{DM}}{m_s}
\approx \frac{10^{-41}}{m_s} \; \gev^4 \,.
\end{align}
Inserting the maximum allowed value from Higgs to invisible searches
for the Higgs portal coupling, $\lambda_{hs}=8.7\cdot10^{-3}$, we
arrive at a mean free path of 
\begin{align}
\lambda
=  \frac{4 \pi}{\lambda_{hs}^2 g_{h\gamma\gamma}^2} \, \frac{m_h^3}{\rho_\text{DM}} \gtrsim   10^{43}\  \text{m}\,.
\end{align}
The crucial observation is that a light DM mass cancels between the
cross section and the particle density.  Hence, the mean free path for
light DM particle ($m_s\ll m_h$) due to scattering at the DM
background is universally bigger than the size of the observable
universe, $l_\text{univ}\approx 30\, \text{Gpc} \approx 10^{27} $~m,
by a factor of $10^{16}$.  That's not good news for the LHC.

For slightly larger DM masses we can briefly look at the competing
annihilation channel
\begin{align}
\langle s \rangle s \to f \bar{f} \,,
\end{align}
with the cross section 
\begin{align}
\sigma_{\langle s \rangle s\to \bar{f}f}
=\frac{\lambda_{hs}^2}{8\pi}\frac{m_f^2}{m_h^4}\left(1-\frac{4m_f^2}{m_s m_h}\right) \; ,
\end{align}
where we again use $s =m_s m_h$ and assume $m_s\ll m_h$.  This cross
section is always positive above threshold, which for electrons is
$m_s > 8.4$~eV, but it decreases for small DM mass. For electrons and
$m_s \sim 10$~eV the relic number density will become very small and
the mean free path will still be $\lambda \approx 10^{39}\ \text{m}$,
and the process hence unobservable.

To summarize, the mean free path of a dark matter particle in the DM
background field with our local DM density
can be written as
\begin{align}
\lambda = 8.5\,\text{m} \; 
\frac{m_s}{10^{-22}\text{eV}} \; 
\frac{10^{-6}~\text{GeV}^{-2}}{\sigma_{\langle s \rangle s\to \text{sth.}}} \; .
\label{eq:lambda_general}
\end{align}
The first term implies that the DM abundance increases with decreasing
DM mass, and the mean free path decreases. The second term says that
higher cross sections also shorten the mean free path. Applied to
fuzzy DM with $m_s \sim 10^{-22}$~eV annihilating to $\gamma\gamma$,
the first term becomes $\ord (1)$ but due to the $m_s/m_h^3$
suppression in the cross section, we cannot even come close to the
$10^{-6}~\text{GeV}^{-2}$ of the numerator. For annihilations into
$e^{+}e^{-}$ the cross section is larger but still suppressed by
$m_e^2/m_h^4$, \ie nowhere close to $10^{-6}~\text{GeV}^{-2}$. In
either case, we face a large Higgs mass suppression coming from the
Higgs propagator in the annihilation process.  The only way out of
this is to introduce a new scale that compensates or replaces at least
some suppression factors. Replacing the light scalar with a
pseudoscalar does not help, either. Instead, it adds a momentum
suppression relative to the Higgs mass which further reduces the rate.

\subsection{Dark matter scattering}
\label{sec:LHC_DD}

\begin{figure}[b!]
\begin{center}
\includegraphics[width=.4\textwidth]{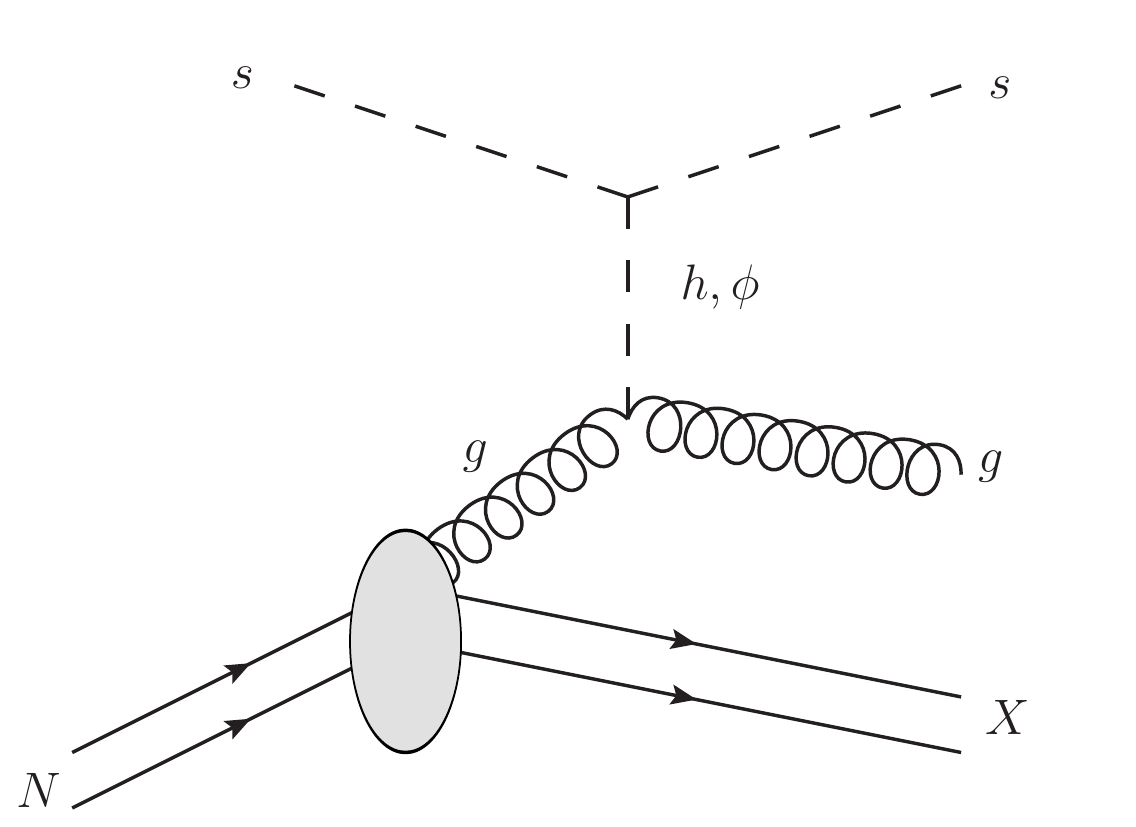}
\end{center}
\caption{Deep inelastic scattering of a light scalar $s$ off a
  nucleon.}
\label{fig:phi_gluglu} 
\end{figure}

The second LHC process we explore is that of a light scalar produced in Higgs
decays scattering with the detector material. A sufficiently light
scalar will be highly energetic and can break up the nucleus in deep
inelastic scattering. This will lead to a characteristic signature of
a spontaneously appearing hadronic jet in the dense calorimeter
material. This signature is inspired by direct detection of dark
matter, even though in our case it would not confirm the dark matter
nature of the light new scalar. The Feynman diagram for the process is
shown in Fig.~\ref{fig:phi_gluglu}. The difference to the usual DIS
process in high energy physics is that the nucleus $N$ is not highly
relativistic. Instead, approximanting $E_s \approx m_h/2$, the
center-of-mass energy of the scalar--nucleus scattering is $\sqrt{s} =
\sqrt{2 E_s \,M} \approx \sqrt{m_h M}$.  As in the last section, we
start with the kinematics of our process before we compute the hard
matrix element. To compute deep inelastic scattering of a scalar $s$
off the detector material of LHC experiments we will once more
consider the mean free path,
\begin{align}
\lambda = \frac{1}{n_\text{det} \; \sigma_\text{DIS}}\,,
\end{align}
of a light scalar $s$ in the detector material with the number density
$n_\text{det}$.  As an example we consider the ATLAS
calorimeters. Both, the electromagnetic (ECAL) and hadronic
calorimeter (HCAL) are sampling detectors using lead and iron as
absorber materials~\cite{Aad:2008zzm}.  As DIS takes place at the
level of nucleons rather than the full nuclei we are interested in the
nucleon density per unit target material. The effective nucleon
density for a material $X$ can be computed as,
\begin{align}
  n_X =
  N_A\ \rho_X\ \frac{A_X}{m_X^\text{mol}}\; ,
\end{align}
where $N_A$ denotes Avogadro's number, $A_X$ the mass number, $\rho_X$
the density, and $m_X^\text{mol}$ the molar mass of the material $X$.
In the central region of the detector the ECAL has a radial extension
of $L_\text{E} = 0.6$~m and the HCAL of $L_\text{H} = 2$~m. The inner
tracking detector is a gas detector and hence can be neglected due to
its low number density.

For each of the two detector materials we can compute the partonic
cross section for the process shown in Fig.~\ref{fig:phi_gluglu},
\begin{align}
s N \to s g + X \; .
\end{align}
The partonic DIS cross section is usually given in terms of the energy
loss $\nu$ of the scalar and the two kinematic variables $x = Q^2/(2 M
\nu)$ and $y= \nu/E_s$.  The nucleon scattering cross section can then
be expressed as the incoherent sum of partonic cross sections weighted
by their respective parton distribution function,
\begin{align}
\frac{d\sigma_\text{DIS}}{dx\, dy} = \sum_j \,\frac{d\hat\sigma_\text{DIS}}{dx\, dy} \ f_j(x, Q^2) \,.
\label{eq:convpdfs}
\end{align}
As long as we only consider DM particles dominantly coupling to
gluons, the incoherent sum in Eq.\eqref{eq:convpdfs} reduces to
weighting of the partonic cross section with the gluon PDF.  To study
DIS in the material of the LHC detectors we need to take into account
nuclear effects by using the nuclear rather than proton PDF
sets. Therefore, we perform all our calculations with the
\texttt{nCTEQ15} nuclear PDF set~\cite{Kovarik:2015cma} via the
\texttt{ManeParse} package~\cite{Clark:2016jgm}. As a cross check we
compare the results with those using the MMHT proton
densities~\cite{Harland-Lang:2014zoa}. For instance in case of a
scalar mediator we find about $40\%$ more events predicted by the
appropriate proton PDF sets.  Once we know the interaction rate
$\sigma_\text{DIS}$ we can compute the probability that a single
particle $s$ scatters in a detector of length $L_\text{det}$ as
\begin{align}
P_\text{DIS} 
= 1- e ^{- \sum L/\lambda} \; .
\end{align}

\paragraph{Scalar Higgs portal}
The differential cross section for the hard scattering process in the
Higgs portal model is given by
\begin{align}
\frac{ d^2 \hat\sigma_\text{DIS}}{dx\, dy} = \frac{\lambda_{hs}^2
  g_{hgg}^2}{4 \pi \, \hat s } \; \frac{Q^4 }{(Q^2 + m_h^2)^2} \; ,
\end{align}
where $\hat s = x s = 2 M E_s\, x$ and $Q^2=2 M E_s\, x\, y$.
The loop-induced Higgs coupling to gluons, $g_{hgg} =
\alpha_s/(12\pi)$, is defined in Eq.\eqref{eq:higgsgluon}. This
partonic cross section has to be convoluted with the gluon density in
the heavy nucleus and integrated over the full phase space. In the
case of the scalar Higgs portal the full DM DIS cross section on lead
and iron evaluate numerically to
\begin{align}
\sigma_\text{Fe} = 5.3\cdot 10^{-9}\, \text{fb} \qquad \text{and}\qquad \sigma_\text{Pb} = 5.5\cdot 10^{-9}\, \text{fb} \,.
\end{align}
The total scattering probability of a particle moving radially
outwards is then given by
\begin{align}
P_\text{DIS} 
= 1- e ^{-L_\text{E}\, n_\text{Pb}\, \sigma_\text{Pb}}  e ^{-L_\text{H}\, n_\text{Fe}\, \sigma_\text{Fe}}  
\approx7.5 \cdot 10^{-21}  \; .
\label{eq:pdis}
\end{align}
We can combine this scattering probability with the Higgs production
rate at the LHC and compute the expected number of dark matter DIS
events for the maximum allowed branching ratio  from supernova
cooling constraints, $\lambda_{h s} \approx 2.5\cdot10^{-4}$,
discussed in Sec.~\ref{app:SN},
\begin{align}
N_\text{DIS}  
= \mathcal{L} _\text{HL} \; \sigma_h \, \text{BR}_{h\to ss} \;  P_\text{DIS}
\approx 4.1 \cdot 10^{-16} \; .
\label{eq:ndis}
\end{align}
Inserting the Higgs production rate at $\sqrt{s}=14$~TeV of around
$\sigma_h \approx 60$~pb~\cite{deFlorian:2016spz} and the total
integrated luminosity expected in the high-luminosity run of the LHC
(HL-LHC) of $\mathcal{L} _\text{HL}\approx 3 \ \text{ab}^{-1}$, we
find that this process is hopeless to observe in the renormalizable
Higgs portal model.

\begin{figure}[t]
\includegraphics[width=.51\textwidth]{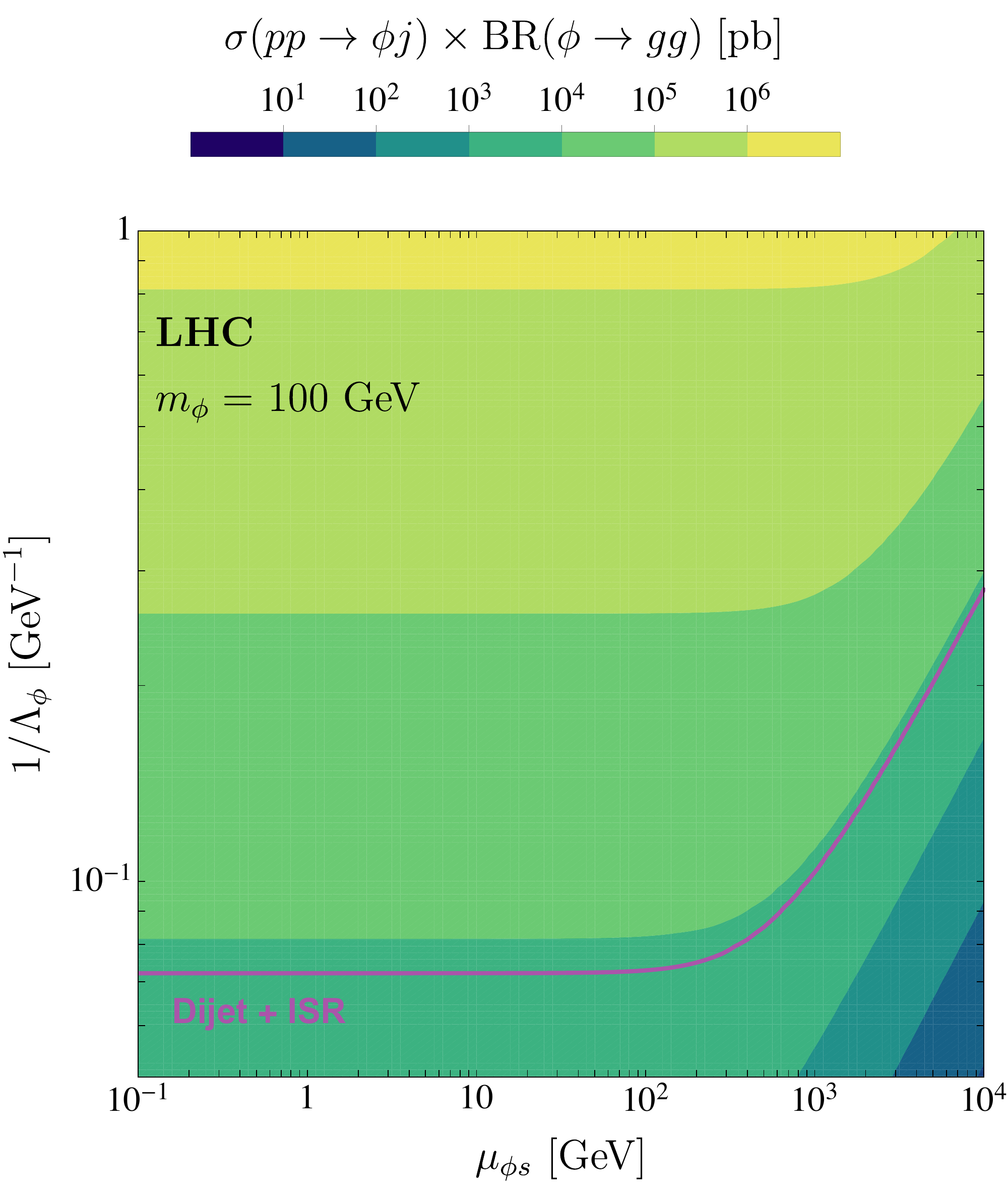}%
\includegraphics[width=.5\textwidth]{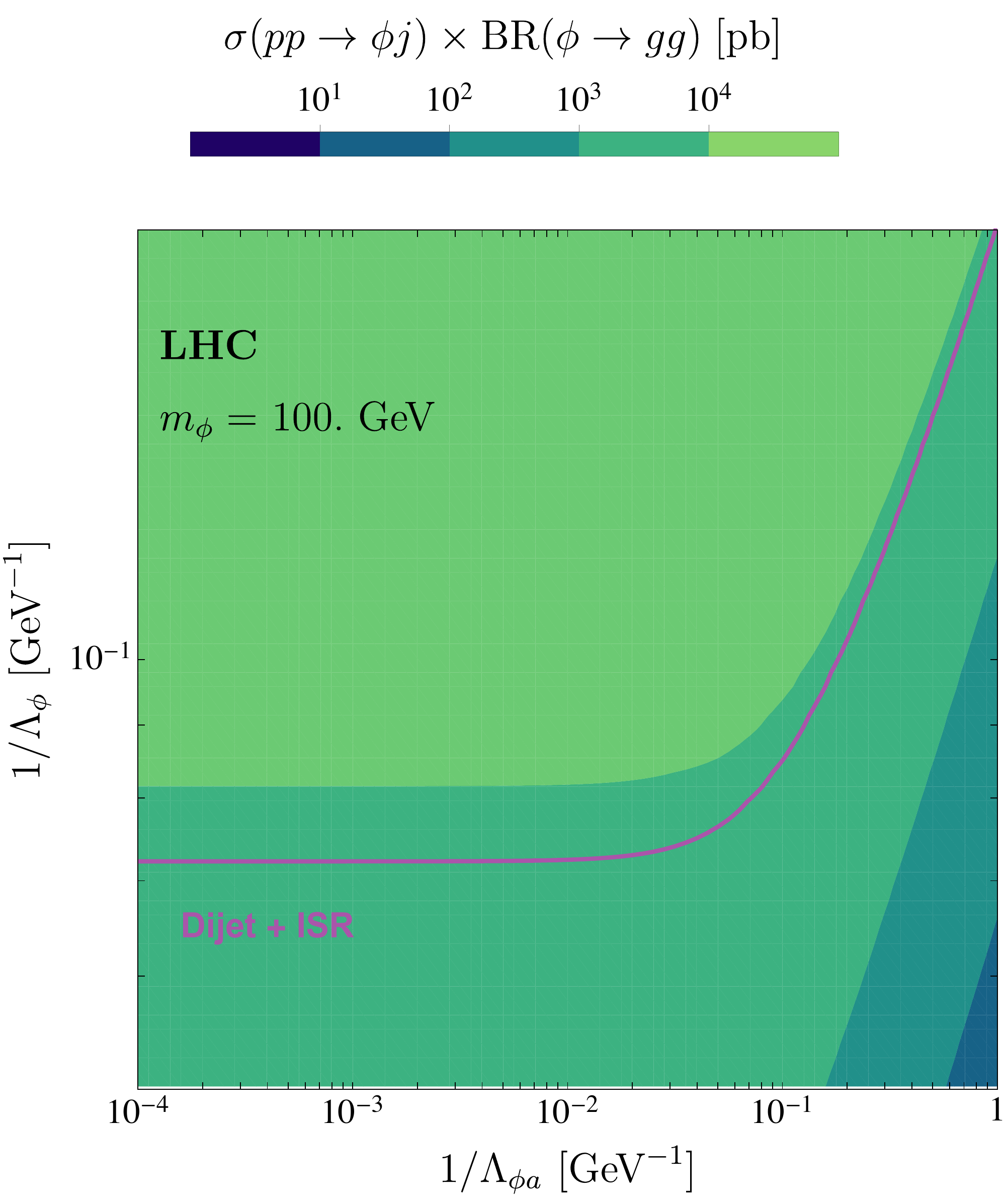}
\caption{Cross section times branching ratio for the process $ p p \to
  (\phi \to g g ) + j$ for $m_\phi =100$~GeV for the scalar (left) and
  pseudoscalar (right) model with a scalar mediator. The purple
  contour shows the CMS cross section limit of
  Ref.~\cite{Sirunyan:2017nvi}.}
\label{fig:xsec_dijet} 
\end{figure}

\paragraph{Scalar with new mediator}
A more flexible alternative to the renormalizable Higgs portal is a
new scalar mediator $\phi$ with an effective coupling to
gluons. Before we study the DIS signature at the LHC we note that such
a mediator can decay into a pair of gluons or a pair of DM particles,
\begin{align}
\Gamma_{\phi\to gg} 
&= \frac{2\, \alpha_s^2}{\pi} \frac{m_\phi^3}{\Lambda_\phi^2}\,, \notag \\
\Gamma_{\phi\to ss} 
&= \frac{1}{32 \pi} \frac{\mu_{\phi s}^2}{m_\phi} \sqrt{1- 4 \frac{m_s^2}{m_\phi^2}}\,.
\end{align}
This means that the coupling to gluons will always lead to a di-jet
resonance $\phi \to g g$. At the LHC this resonance can be searched
for in a bump hunt on top of a smoothly falling QCD di-jet background.

Even though the corresponding limits cannot be simply translated from
narrow resonance searches to our model with a potentially wide
mediator, we briefly report on the rough order of existing
constraints.  In the low-mass regime of $50 - 300$~GeV, CMS has looked
for vector resonances $Z'\to \bar q q$ in di-jet events with an
additional jet from initial state radiation
(ISR)~\cite{Sirunyan:2017nvi,Sirunyan:2019vxa}. In this search, the
di-jet system from the $Z'\to \bar q q $ resonance has to recoil
against a hard ISR jet, where at least one of the jets satisfies $p_T
> 500 $~GeV. This cut ensures that the low-mass $Z'$ is heavily
boosted and the di-jet from the resonance will be reconstructed as a
single large jet, back to back with the ISR jet.  We can reinterpret
this analysis for our mediator model in a simplistic way by generating
the $(\phi \to g g ) + j$ signal with just a cut of $p_T > 500$~GeV on
the scalar.  We generate the relevant signal events, $p p \to \phi +
j$, with~\texttt{MadGraph5\_aMC@NLO}~\cite{Alwall:2014hca}.  The
corresponding cross section times branching ratio is shown in
Fig.~\ref{fig:xsec_dijet}. The purple contour represents the CMS cross
section limit for a 100~GeV mediator derived in
Ref.~\cite{Sirunyan:2017nvi}. Such a boosted jet analysis should be
more stable for a broadening resonance than for instance a
trigger-level resonance search.  Nevertheless, we do not claim that an
actual analysis for our model will ever be as good as the narrow-width
$Z'$ search and only quote the CMS limits as the most optimistic
estimate.

\begin{figure}[t]
\includegraphics[width=.5\textwidth]{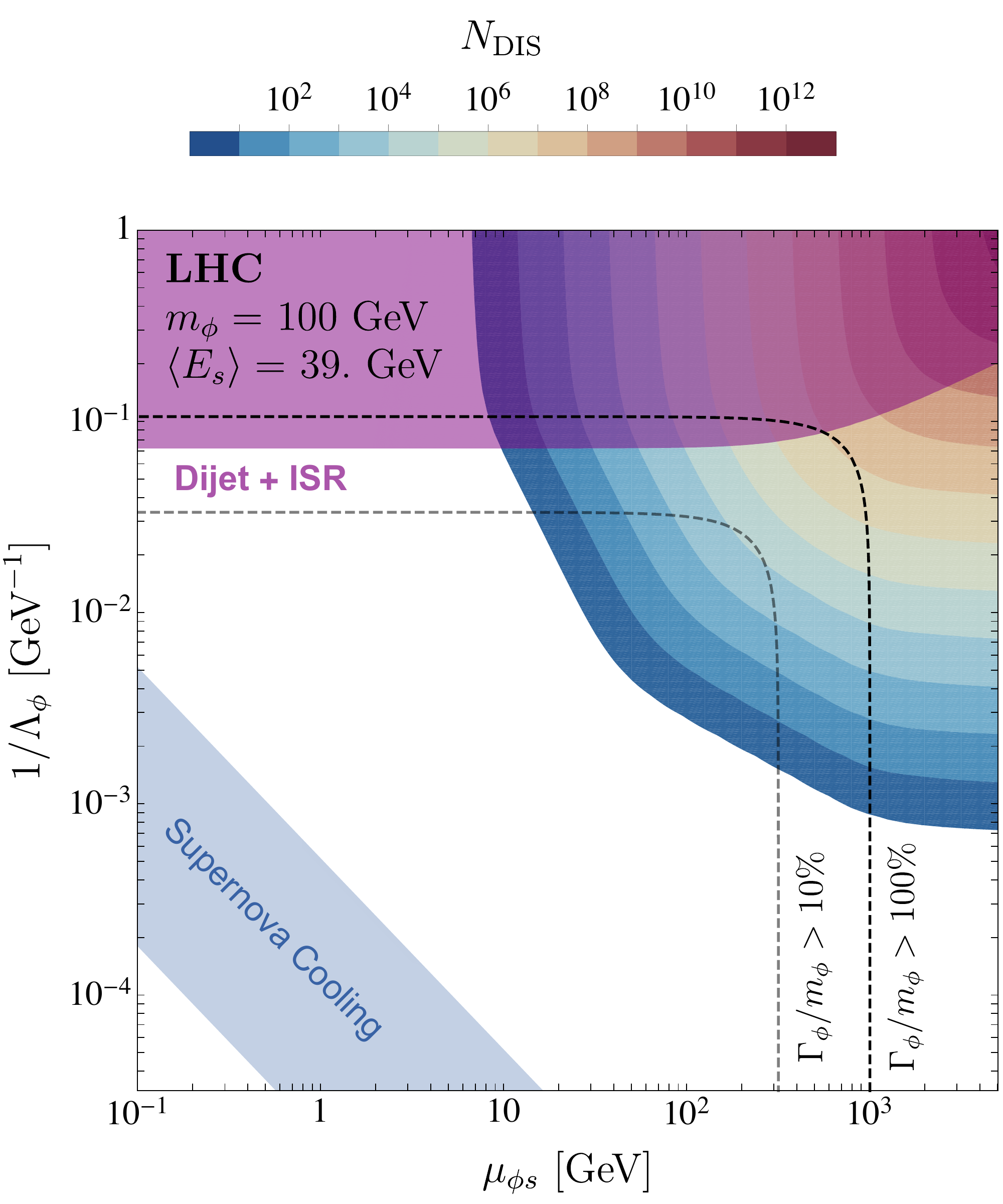}
\includegraphics[width=.51\textwidth]{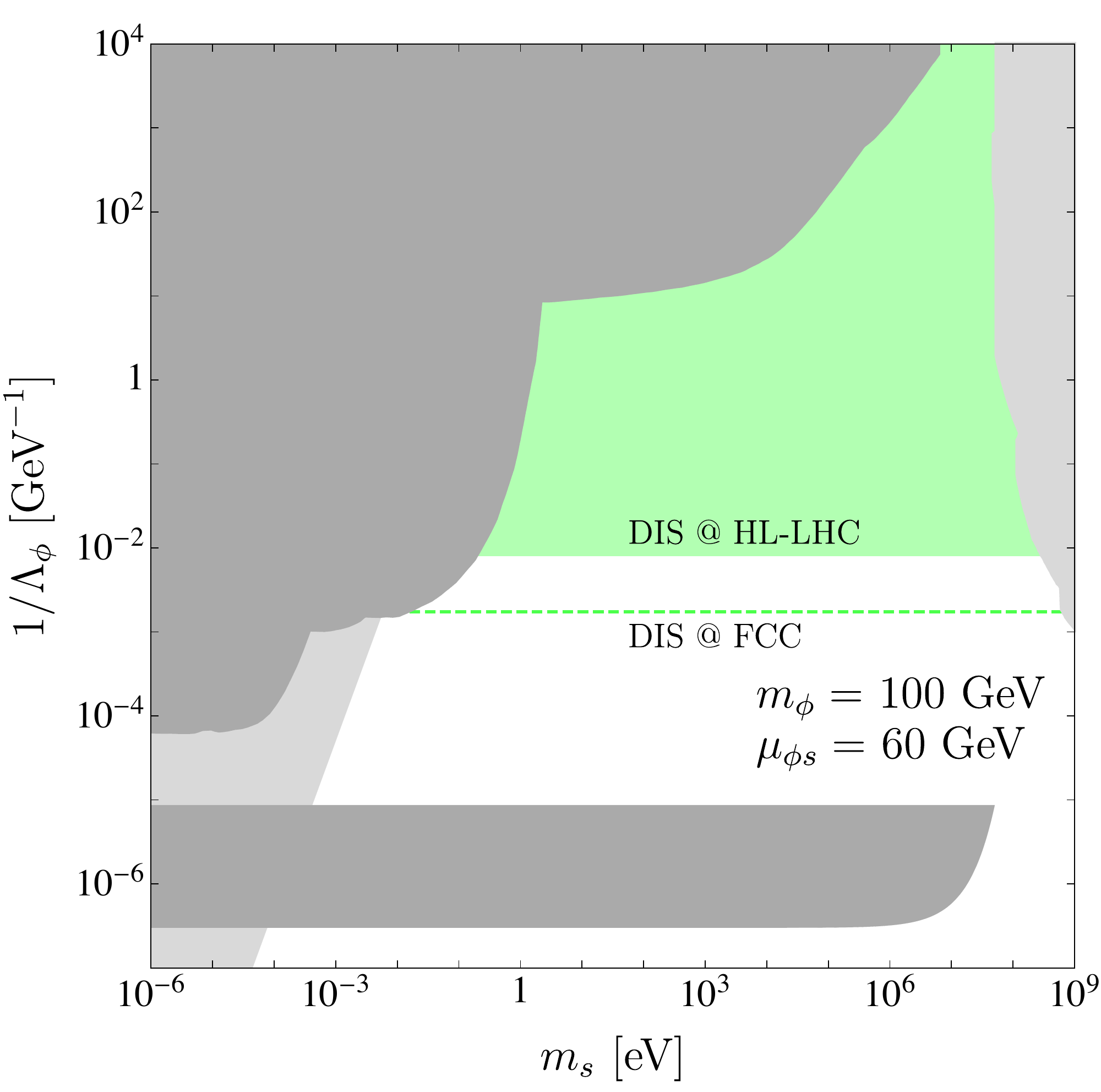}
\caption{Left: number of expected DIS events in the plane of
  DM-mediator coupling $\mu_{\phi S}$ versus mediator-gluon coupling
  $\Lambda_{\phi}$ for $m_\phi=100$~GeV.  The blue band represents the
  constraint from Supernova cooling, the purple area is the bound on
  low-mass di-jet resonances~\cite{Sirunyan:2017nvi}, and the dashed
  lines indicate fixed ratios $\mu_{\phi S}/\Lambda_\phi$. Right:
  comparison of the projected DIS reach with the low-energy
  constraints in terms of $\Lambda_\phi$ for fixed $\mu_{\phi
    s}$. Constraints which require the dark matter nature are shown in
  light grey.}
\label{fig:NdisSSimp} 
\end{figure}

In further analogy to the Higgs portal case, the DM scalars $s$ can
undergo DIS in the detector material via a mediator $\phi$. We can
again calculate the partonic DIS cross section,
\begin{align}\label{eq:disSSimp}
  \frac{d^2\hat\sigma_\text{DIS}}{dx\, dy}
  = \frac{\alpha_s^2}{4 \pi\, \hat s}\,
  \left( \frac{\mu_{\phi s}}{\Lambda_\phi} \right)^2 \; \frac{Q^4 }{(Q^2 + m_\phi^2)^2}  \; .
\end{align}
To compute the total number of expected DIS events we simulate DM
production, $p p \to \phi \to s s$, with
\texttt{MadGraph5\_aMC@NLO}~\cite{Alwall:2014hca} for a large range of
mediator couplings to gluons $(1/\Lambda_\phi)$ and couplings to the
dark matter scalar ($\mu_{\phi s}$). The corresponding number of
expected DIS events at the HL-LHC is shown in the left panel of
Fig.~\ref{fig:NdisSSimp} for a mediator of mass $m_\phi =100$~GeV.
The purple area is approximately excluded by the di-jet limit. In our
Monte Carlo study we analyze the $p_T$- spectra of the produced DM
particles $s$ and confirm that in the regime of a narrow mediator,
$\Gamma_\phi/m_\phi\lesssim 10\%$, the averaged DM energy is $\langle
E_s\rangle \approx 39$~GeV, with the bulk of the particles having
$E_s=m_\phi/2 = 50$~GeV. To compute the number of DIS events we assume
this average energy over all displayed parameter space. This is
conservative, as in the case of a broader resonance the produced DM
particles become more energetic on average, thus enhancing the DIS
cross section. In a fully-fletched analysis the scattering cross
section should be convoluted with the DM energy spectrum. However,
this is beyond the scope of this sensitivity study. Similarly, we
assume that the displaced recoil jet signature is essentially
background-free. At the LHC this statement is never strictly true,
because for instance detector failures or support structures can of
course generate displaced objects. Moreover, if a highly energetic jet
were to consist only of long-lived neutral hadrons it could generate such a
recoil, but such a strong suppression of all charged hadrons is
rather unlikely.

As always, passing the LHC triggers is the first challenge for our
signal. Barring other trigger opportunities we can rely on the
standard mono-jets trigger requiring missing transverse momentum
around 100~GeV. We emphasize that in addition to the standard
mono-jets signature we can use the displaced recoil jet to reduce the
SM-backgrounds. On the other hand, there might be additional handles
on the trigger, so we will quote projected limits without the
trigger requirements. We have checked that a for a
momentum-independent scalar coupling the trigger catches at least 10\%
of the signal rate and weakens the projected limits in the coupling
$1/\Lambda_\phi$ by at most a factor three.

In the right panel of Fig.~\ref{fig:NdisSSimp} we contrast the
projected reach of the DIS process with the low-energy and other
limits from Fig.~\ref{fig:scalar-mediator}. We see that the DIS probe
is complementary to all other constraints and fills the gap for dark
matter masses between 1~eV and 100~MeV over two orders of magnitude in
the gluon coupling $1/\Lambda_\phi$. The projected FCC sensitivity
corresponds to a collider energy of 100~TeV and a luminosity of $\lumi
=30 \ \text{ab}^{-1}$~\cite{Benedikt:2018csr}. More details on the FCC
estimate can be found in App.~\ref{app:fcc}. In terms of parameter
reach the FCC projections exceed the HL-LHC projections by another
order of magnitude in the coupling.

\paragraph{Pseudoscalar with Higgs mediator}
The third model we consider for the DIS signature is ALP dark matter
with a derivative Higgs portal.  Repeating the previous calculation
for the analogous process with the scalar $s$ replaced by the
shift-symmetric scalar $a$ yields the differential DIS cross section
\begin{align}
\frac{ d^2 \hat \sigma_\text{DIS}}{dx\, dy}=  \frac{g_{hgg}^2 }{16 \pi \, \hat s } \ \frac{ Q^4}{\Lambda_{ha}^4} \left(\frac{Q^2 +2 m_a^2}{Q^2 + m_h^2}\right)^2 \; ,
\end{align}
where $\hat s = x s = 2 M E_a\, x$ and $Q^2=2 M E_a\, x\, y$.  We
always assume the minimum suppression scale from the Higgs to
invisible limit of $\Lambda_{a h} \approx 832$~GeV.  Integrating this
cross section in the limit $m_a \ll m_h$ and using once more the ATLAS
calorimeter materials and dimensions as a benchmark, the full DM DIS
cross section on lead and iron give us
\begin{align}
\sigma_\text{Pb} = 5.7\cdot 10^{-12}\, \text{fb} \qquad \text{and}\qquad \sigma_\text{Fe} = 6.5\cdot 10^{-12}\, \text{fb} \,.
\end{align}
This means that a single produced pseudoscalar $a$ undergoes DIS in
the detector with a probability of
\begin{align}
P_\text{DIS}  \approx8.8 \cdot 10^{-24} \,. 
\end{align}
The total number
of expected DIS events is given by
\begin{align}
N_\text{DIS} \approx 2.2 \cdot 10^{-16} \,.
\end{align}
As for the ULDM scalar with the Higgs portal, this process is
unobservable at the HL-LHC.

\begin{figure}[t]
\includegraphics[width=.5\textwidth]{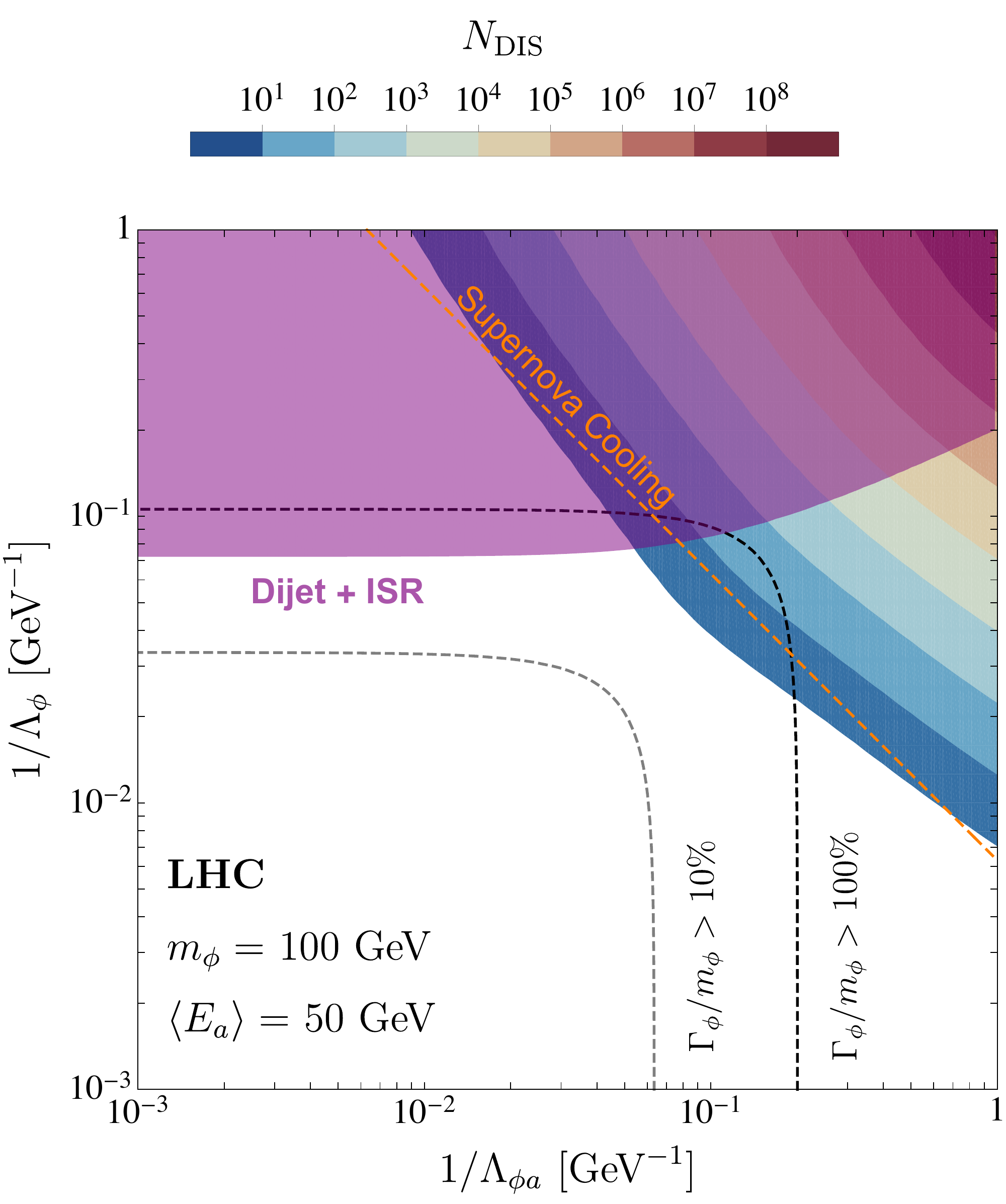}
\includegraphics[width=.51\textwidth]{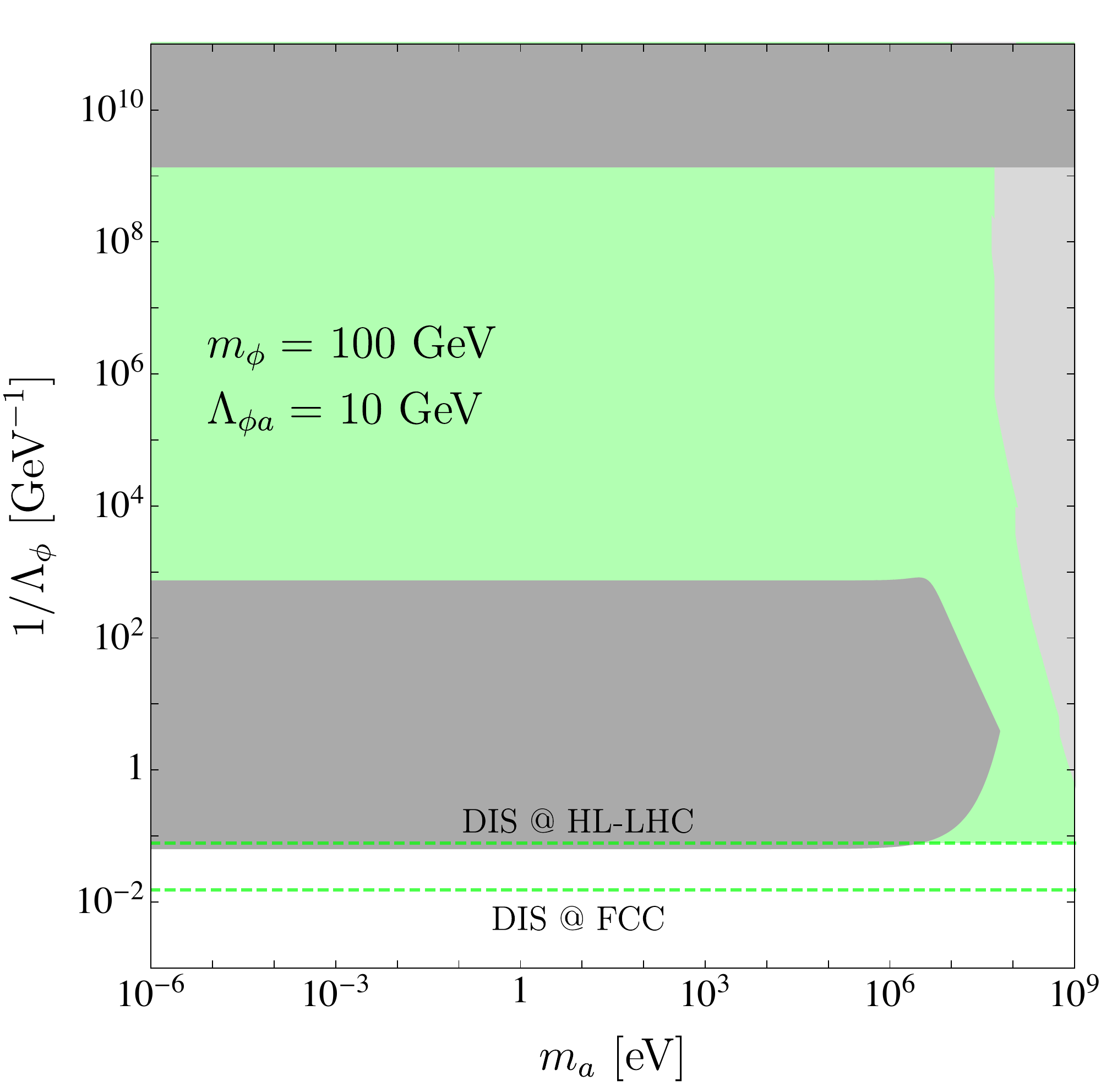}
\caption{Left: number of expected DIS events in the plane of
  DM-mediator coupling $\Lambda_{\phi a}$ versus mediator-gluon
  coupling $\Lambda_\phi$ for $m_\phi =100$~GeV. Right: comparison of
  the projected DIS reach with the low energy constraints in terms of
  $\Lambda_\phi$ for fixed $\Lambda_{\phi a} =10$ GeV. Constraints
  which require the dark matter nature are shown in light grey.}
\label{fig:NdisDSimp} 
\end{figure}

\paragraph{Pseudoscalar with new mediator}
In the same spirit as for the scalar case, we again consider the
simplified model for the pseudoscalar $a$ with a new scalar mediator.
As the shift-symmetric pseudoscalar couples only via derivatives, its
interaction strengths are in general momentum-dependent. At the LHC
$a$ is produced from the decaying mediator $\phi$, which again is
produced in gluon fusion. As in the scalar case, we generate the
corresponding signal, $p p \to \phi \to a a$, with
\texttt{MadGraph5\_aMC@NLO}~\cite{Alwall:2014hca}. Our Monte Carlo
study shows that as long as the mediator has a comparatively narrow
width, $\Gamma_\phi/m_\phi\lesssim 10 \%$, the produced DM particle
$a$ carries roughly half the mediator mass in momentum. For a
weak-scale mediator mass $m_\phi\sim v$ this is typically enough such
that $a$ can undergo deep inelastic scattering with the nuclei in the
detector material. The relevant cross section of the hard scattering
process,
\begin{align}
  a N \to a g + X\,,
\end{align}
reads
\begin{align}\label{eq:disDSimp}
\frac{d^2\hat \sigma_\text{DIS}}{dx\, dy} = \frac{\alpha_s^2}{16\pi\, \hat s}\,\frac{Q^4}{\Lambda_{\phi a}^2\Lambda_\phi^2} \, \left( \frac{Q^2+2m_a^2}{Q^2+m_\phi^2}\right)^2\,.
\end{align}
The left panel of Fig.~\ref{fig:NdisDSimp} displays the expected
number of DIS events at an ATLAS-like detector for the high luminosity
run of LHC. The grey and black dashed lines show the contours of
$\Gamma_\phi/m_\phi\lesssim 10 \%$ and $\lesssim 100 \%$. The effect
of the mono-jets trigger is significantly smaller than for the scalar
case, because of the momentum-dependent DM-coupling. We estimate the
trigger survival rate of the pseudoscalar signal to be above 70\%,
translating into a negligible 15\% shift in the coupling reach.

Again, the right panel of Fig.~\ref{fig:NdisDSimp} shows the
corresponding HL-LHC and FCC projections in model space, compared to
all other limits from Fig.~\ref{fig:alp_odd}. The DIS signature closes
the wide gap from all other limits over 13 orders of magnitude in
$m_a$ and covers the supernova constraints, providing an independent
collider probe of the cosmological observations. The FCC projections
again exceed the HL-LHC projections by an order of magnitude in the
coupling and provides the leading signatures for pseudoscalar dark
matter with a weak-scale mediator.

\section{Conclusions}
\label{sec:conclusions}

Light dark matter is a relatively new avenue of dark matter model
building and phenomenology. It leads us to consider a wealth of new
measurements and cosmological observations, while challenging
established search strategies like large-scale direct detection of
hadron collider analyses. It also forces us to go back and forth
between a semi-classical wave-like description and a quantum field
theory Lagrangian. 

We have studied models with a light scalar or pseudoscalar dark matter
agent with a mass ranging from well below the eV scale to the GeV
scale with a focus on bridging the quasi-classical and quantum
descriptions. As mediators to the Standard Model we have assumed the
SM-like Higgs or a new, weak-scale scalar. We have studied a large
number of constraints from low-energy precision measurements, big bang
nucleosynthesis, supernova cooling, invisible Higgs decays, and direct
dark matter detection. The relative impact of these constraints
depends strongly on the quantum numbers of the dark matter and on the
nature of the mediator. While BBN strongly constrains very light
scalar dark matter and invisible Higgs decays obviously only apply to
models with a SM-like Higgs mediator, supernova constraints are very
model independent. Current direct detection experiments, like
Xenon1T, start cutting into the parameter space at relatively
large dark matter masses. The difference between scalar and
pseudoscalar dark matter can largely be understood by the derivative
interactions of the pseudoscalar and its effect on the interaction
rates.

Inspired by the process topologies of direct and indirect detection we
have studied two novel LHC signatures. On the one hand, light dark
matter particles produced for instance in Higgs decay can annihilate
with the DM background in the LHC detector. In analogy to indirect
detection, but with the benefit of probing the actual dark matter
property, we can look for pairs of photons or electrons produced in
the LHC detectors. Unfortunately, we find the rate of this signal to
be very low. On the other hand, similar to direct detection in nucleon
recoils, light dark matter produced at the LHC can hit the nuclei in
the ATLAS and CMS calorimeters and produce a \textsl{hard, displaced
  recoil jet}. This completely new signature of ULDM should be
observable for scalar mediator models and over the full range of dark
matter masses below the GeV scale. For scalar ULDM this signature
closes a gap in all current constraints from $m_s = 1$~eV to the
direct detection thresholds around $m_s = 100$~MeV. For pseudoscalar
ULDM it closes the entire gap between atomic spectroscopy measurements
and supernova cooling for $m_a = 10^{-6}~...~10^7$~eV, again all the
way to large-scale direct detection experiments. Unlike direct
detection signals this LHC search does not assume a relic density, and
the field-theoretical description allows for a consistent comparison
to cosmological supernova observations.  In that sense it can also be
immediately generalized to other very light particles, for instance
neutrinos, as long as they can be produced at the LHC with sufficient
energy and rate.

\begin{center} \textbf{Acknowledgments} \end{center}

We thank Joerg Jaeckel, Felix Kling, Simon Knapen, Ennio Salvioni and Hans-Christian Schultz-Coulon for useful discussions.
MB and PF are funded by the UK Science and Technology Facilities Council (STFC) under grant ST/P001246/1. 
TP is supported by the DFG Transregio \textsl{Particle physics
  phenomenology after the Higgs discovery} (TRR~257).  PR is funded by
the DFG Graduiertenkolleg \textsl{Particle physics beyond the Standard
  Model} (GRK~1940).

\clearpage
\appendix
\section{Calculational details}

In this Appendix we collect details of the various computations performed to derive the various limits in this paper.

\subsection{Variation of fundamental constants}
\label{app:fundamental}

Measurements that are sensitive to variations in the fundamental
constants $m_f, \alpha$ and $m_V$ set stringent constraints on models
including scalars and pseudoscalars for ULDM with masses below
$m_s\ll$~eV. SM-like operators that describe the fermion masses $m_f$,
the coupling $\alpha$, and the gauge boson masses $m_V$ are
\begin{align}
\lag_\text{SM} \supset  -\sum_f m_f \bar{f}f -
\frac{F_{\mu\nu}F^{\mu\nu}}{4}+\sum_V\delta_Vm_V^2V_\mu V^\mu \; .
\end{align}
with $\delta_W=1$ and $\delta_Z=1/2$.  For instance in the Higgs
portal model of Eq.\eqref{eq:higgsportal} these constants are modified
by the effective interaction operators
\begin{align}
\lag \supset
\frac{\lambda_{h s}}{2} \frac{m_f}{m_h^2} \; s^2 \bar{f}f
- \frac{\lambda_{h s} g_{h\gamma\gamma}}{2} \frac{1}{m_h^2} \; s^2 F_{\mu\nu}F^{\mu\nu}
- \lambda_{h s} \delta_V \frac{m_V^2}{m_h^2} \; s^2  V_\mu V^\mu \; .
\end{align}
For the scalar model of Eq.\eqref{eq:derivativePortal} with a new
mediator the corresponding effective Lagrangian reads
\begin{align}
  \lag \supset
  - \frac{m_a^2 m_f}{2 \Lambda_{ha}^2 m_h^2} \; a^2 \bar{f}f
  + \frac{g_{h\gamma\gamma}}{2} \frac{m_a^2}{\Lambda_{ha}^2 m_h^2} \; a^2 F_{\mu\nu}F^{\mu\nu}
  + \delta_V \frac{m_a^2  m_V^2}{\Lambda_{ha}^2m_h^2} \; a^2  V_\mu V^\mu~.
\end{align}
Wherever we treat the DM field as a classical field, we can simply
translate these two sets of Lagrangians into each other via
$\lambda_{h s}s^2 \leftrightarrow - m_a^2/\Lambda^2 a^2$.  In both the
scalar and pseudoscalar case, we insert the classical field solution
and rewrite the quadratic (pseudo-) scalar interaction in a constant
and a time-dependent part as done in Ref.~\cite{Stadnik:2015kia}
\begin{align}
s^2&= s_0^2 \cos^2(m_st) \to \frac{s_0^2}{2} \left( 1+\cos(2m_st) \right)\notag \\
a^2&= a_0^2 \cos^2(m_at) \to \frac{a_0^2}{2} \left( 1+\cos(2m_at) \right)~.
\end{align}
In this form the constant term describes a fifth force while the
oscillating terms lead to a variation of fundamental constants, for
instance the fermion mass
\begin{align}
m_f  \to m_f \left[
      1+\frac{s^2}{\Lambda_{s,f}^2}\right] = m_f \left[
      1+\frac{s_0^2}{2\Lambda_{s,f}^2}+\frac{s_0^2}{2\Lambda_{s,f}^2}\cos(2m_st)\right]
\end{align}
with $1/\Lambda_{s,f}^2=\lambda_{hs}/(2m_h^2)$.
The variation of the fine structure
constant and the weak boson masses can be derived in complete analogy
with $1/\Lambda_{\gamma,s}^2=2\lambda_{hs}g_{h\gamma\gamma}/m_h^2$ and
$1/\Lambda_{s,V}^2=\lambda_{hs}/m_h^2$ and accordingly for the derivative case.

\subsection{Multi-pseudoscalar exchange}
\label{app:models_multi}

We mention in the main text that a pseudoscalar field $a$ obeying a
shift symmetry can be coupled to SM fields via derivative
couplings. For long-range forces mediated by the exchange of $a$ there are important differences between linear couplings of $a$ to the SM and theories in which only operators with multiple $a$ insertions feature.  
For a linear derivative coupling of $a$ to an axial current,
\begin{align}
\frac{\partial_\mu a}{2 f}\  \bar \psi \gamma^\mu \gamma^5 \psi + \frac{\partial_\mu a}{2 f}\  \bar \chi \gamma^\mu \gamma^5 \chi \; ,
\end{align}
long-range forces between the 
fermions $\psi$ and $\chi$ can be induced due to $s$-channel pseudoscalar exchange.
The corresponding amplitude reads
\begin{align}
\overline{|\mat|^2} &= 
\frac{1}{64 f^4} \frac{1}{(q^2-m_a^2)^2} \; 
\Tr[(\slashed p_2 -m_\psi)\slashed q \gamma^5 (\slashed p_1 +m_\psi)\slashed q \gamma^5] \; \Tr[(\slashed p_3 +m_\chi)\slashed q \gamma^5 (\slashed p_4 -m_\chi)\slashed q \gamma^5] 
\notag \\
& = \frac{ m_\psi^2 m_\chi^2}{f^4\left(1-2 \frac{ m_a^2}{q^2} + \frac{m_a^4}{q^4}\right)} =  \frac{m_\psi^2m_\chi^2}{f^4}+\mathcal{O}\left(\frac{m_a^2}{q^2}\right)\,.
\end{align}
In the limit of large momentum transfer $q\gg m_a$ the leading term of the
amplitude is $q$-independent.

\begin{figure}[t]
\centering
\includegraphics[width=0.57\textwidth]{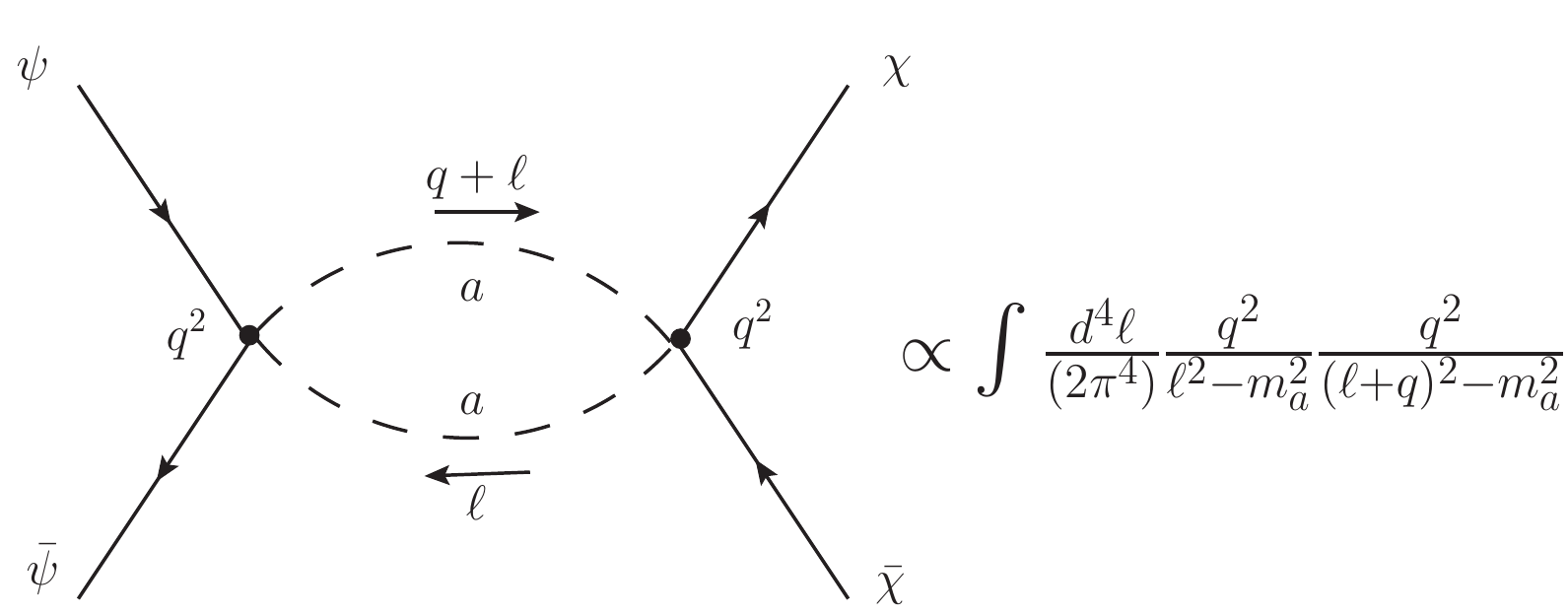}
\hspace*{0.05\textwidth}
\includegraphics[width=0.35\textwidth]{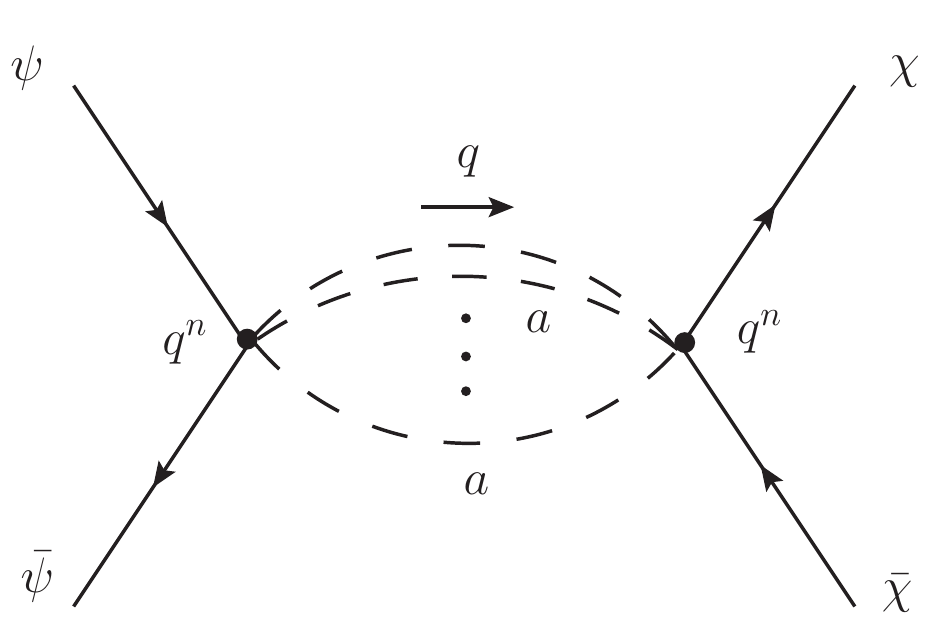}
\caption{Left: Two-to-two scattering via two-pseudoscalar exchange. Right:
  Two-to-two scattering via $n$-pseudoscalar exchange.}
\label{fig:axion_deriv} 
\end{figure}

The leading operator for derivative couplings with two $a$ insertions is given by \eqref{eq:derivativePortal},
\begin{align}
\frac{\partial_\mu a \,\partial^\mu a}{2\Lambda_{ha}^2}H^\dagger H\,.\nonumber
\end{align}
At low energy we can integrate out the Higgs such that the
derivative Higgs portal induces interactions of the type
\begin{align}
\lag \supset \frac{1}{v} \frac{\partial_\mu a \partial^\mu a}{2 \Lambda_{ha}^2} \bar \psi \psi\,+ \frac{1}{v} \frac{\partial_\mu a \partial^\mu a}{2 \Lambda_{ha}^2} \bar \chi \chi\,.
\end{align}
Similar to the case of linear derivative interactions this leads to
the scattering shown in Fig.~\ref{fig:axion_deriv}. The corresponding
matrix element reads
\begin{align}
i \mat &= \bar v (p_2) \frac{i q^2}{2v \Lambda_{ha}^2} u(p_1)\ \int \frac{d^4\ell}{(2\pi^4)}\frac{i}{\ell^2-m_a^2}  \frac{i}{(\ell+q)^2-m_a^2} \ \bar u (p_3) \frac{i  q^2}{2v \Lambda_{ha}^2} v(p_4) \notag \\
&\propto q^4 \int dx  \log\left( \frac{x \Lambda_c}{m_a^2 - x(1-x)q^2}\right)\,,
\end{align}
where $\Lambda_c$ is a momentum cutoff of the loop
integral. Obviously, the resulting amplitude $\overline{|\mat|^2}$ has
no $q$-independent part and vanishes $\propto q^4$ at low momentum
transfer.  Operators with additional derivatives increase the power of
the momentum $q$ associated with the $n$-point vertex. In contrast,
the overall momentum transfer flowing through the diagram will always
be equal to $q^2$. Any long-range force mediated in such a theory is
momentum-suppressed.

Theories with multi-pseudoscalar exchange are therefore qualitatively
different from theories with a linearly coupled pseudoscalar. The
sensitivity of experiments with small momentum exchange is strongly
suppressed in the case of multi-pseudoscalar exchange which makes the
case for complementary approaches beyond astrophysical and precision
measurements of low-energy observables.

\subsection{Big bang nucleosynthesis}
\label{app:bbn}

In the derivation of the BBN limits in Section~\ref{sec:models_scalar} and Section~\ref{sec:models_axion} 
we express the helium yield as in \eqref{eq:deltaY} and ~\cite{Stadnik:2015kia}
\begin{align}
\frac{\Delta (n/p)_W}{(n/p)_W}&=-0.13\frac{\Delta
                                    \alpha}{\alpha}-2.7\frac{\Delta
                                    (m_d-m_u)}{(m_d-m_u)}-5.7\frac{\Delta
                                    M_W}{M_W}+8.0\frac{\Delta
                                    M_Z}{M_Z}\,,\nonumber\\
\frac{\Delta \Gamma_n}{\Gamma_n}&=-1.9\frac{\Delta
                                  \alpha}{\alpha}+10\frac{\Delta
                                  (m_d-m_u)}{(m_d-m_u)}-1.5\frac{\Delta
                                  m_e}{m_e}+10\frac{\Delta
                                  M_W}{M_W}-14\frac{\Delta
                                  M_Z}{M_Z}~.
\label{eq:BBNparam_var}
\end{align}
In the case of a simplified model where only couplings to gluons are present,  Eq.~\eqref{eq:deltaY} becomes particularly simple and one can write~ \cite{Sorensen:2638465}
\begin{align}
\frac{\Delta Y}{Y}=\left(-\frac{Q_{np}}{T_W}+t_\text{BBN}\frac{\partial_x P(x)}{P(x)}\right) \frac{\Delta Q_{np}}{Q_{np}}\   \approx 4.82 \frac{\Delta Q_{np}}{Q_{np}}\,,
\end{align}
where $P(x)$ is the phase space in the neutron decay width~\cite{Sorensen:2638465}.
The energy density of a non-relativistic oscillating DM field is given
by $\rho \simeq m_{s}^2\langle s^2\rangle$ and evolves according to
\begin{align}
\bar{\rho}_{DM}=1.3 \cdot
10^{-6}[1+z(t)]^3~\frac{\text{GeV}}{\text{cm}^3}\,,
\label{eq:oscillating}
\end{align}
with the redshift parameter $z(t)$. For a non-oscillating DM field, we
have $\rho \simeq m_s^2\langle s^2\rangle /2$ and
\begin{align}
\bar{\rho}_{DM}=1.3 \cdot
10^{-6}[1+z(t_m)]^3~\frac{\text{GeV}}{\text{cm}^3}\,,
\label{eq:nonoscillating}
\end{align}
with $z(t_m)$ defined by $H(t_m)\approx m_s$.
In both cases, we assume that the mean DM energy density during weak
freeze-out is much greater than the present-day local cold DM energy density $\langle \phi^2\rangle_{\text{weak}} \gg
\langle \phi^2\rangle_0$. In the case of the oscillating field, we
make use of the relation
$[1+z(t_m)]/(1+z_{\text{weak}})=\sqrt{t_{\text{weak}}/t_m}$ and take
\begin{align}
t_{\text{weak}} &\approx 1.1~\text{s} \nonumber \\
z_{\text{weak}} &\approx 3.2 \cdot 10^9\nonumber \\
H(t_m)&\sim 1/(2t_m) \sim m_s \to t_m \sim 1/(2m_s)
\end{align}
from~\cite{Stadnik:2015kia}.

\subsection{Supernova energy loss}
\label{app:SN}

Stars can be used as a particle-physics laboratory by studying the
energy-loss rate implied by new low-mass particles such as ULDM
particles.  Any annihilation process from SM to light new particles
contributes to supernova cooling. The main assumption in the
corresponding calculation is that the produced particles can freely
escape the supernova, the so-called free-streaming limit. It allows us
to set an upper bound on the coupling strength of the additional
processes. New particles only cause significant effects if they can
compete with the cooling from neutrinos already carrying away energy
directly from the interior of stars. The strongest bound comes from
the SN1987A~\cite{Raffelt:1996wa},
\begin{align}
\varepsilon_x < 10^{19}~\text{erg g}^{-1}\text{s}^{-1}~.
\end{align}
To set constraints, one has to evaluate the novel energy-loss rates at
typical core conditions with a temperature and density of around
\begin{align}
  T_\text{SN}&=30~\text{MeV}
  \qquad \text{and} \qquad 
\rho_\text{core} =3 \cdot 10^{14} ~\text{g cm}^{-3}~.
\label{eq:coreval}
\end{align}
or directly set limits on the total energy-loss rate per unit mass,
\begin{align}
\Gamma = \varepsilon_x \rho_{\text{core}} < 10^{-14} ~\text{MeV}^5~.
\label{eq:SNlimit}
\end{align}

After being produced, new particles travel through the SN core and
might start interacting with the supernova. The process
considered here is elastic scattering $ N s \to N s$ with a
$t$-channel Higgs or scalar particle exchange. To estimate for which
couplings light scalars start to interact with the supernova
particles, we compare the radius of the supernova, $R_\text{SN}
\approx10$~km, with the mean free path of elastic scattering,
\begin{align}
\lambda=\frac{1}{n_N(r)\,\sigma_{sN\to sN}}, \quad
  n_N(r)=\begin{cases} \dfrac{\rho_\text{core}}{m_p} & \text{for}~r\leq
    R_\text{SN}\,,\\[5mm] \dfrac{\rho_\text{core}}{m_p} \left( \dfrac{R_\text{SN}}{r} \right)^m & \text{for}~r>R_\text{SN}\,,\end{cases}
\end{align}
and $m=3~...~7$ depending on the profile chosen.  This condition
characterizes the trapping regime, \ie the point where the new scalars
start thermalizing and are trapped sucht that they cannot escape the supernova
freely anymore. Once the scalars are trapped
they create a scalarsphere similar to the
axiosphere~\cite{Turner:1987by,Raffelt:1996wa}. In the regime where
the free-streaming limit doesn't apply anymore, the sphere still
looses energy via black-body-radiation. The radius of the sphere $r_0$
can be determined by
\begin{align}
  4\pi r_0^2 \; \frac{g\pi^2}{120} \; T(r_0)^4 < 10^{53}~\text{erg}\text{s}^{-1}\,,
\end{align}
with a temperature profile of $T(r)=T_\text{SN}
(R/r)^{m/3}$~\cite{Zhang:2014wra} and $g=1$ the number of effective
degrees of freedom. It varies between
\begin{align}
r_0=1.7~...~7.2\cdot 10^6~\text{cm}\,,
\label{eq:SphereRad}
\end{align}
depending on $m$. The second condition for the black-body radiation of
a scalarsphere with radius $r_0$ is an upper bound on the optical
depth
\begin{align}
  \int_{r_0}^{\infty}\lambda^{-1}\text{d}r \leq \frac{2}{3}~.
\label{eq:depth}
\end{align}
Combining the optical depth criterion with the upper bound on the
luminosity of the scalarsphere, we will set a bound on the couplings
in our models.

\begin{figure}[b!]
\centering
\includegraphics[width=.4\textwidth]{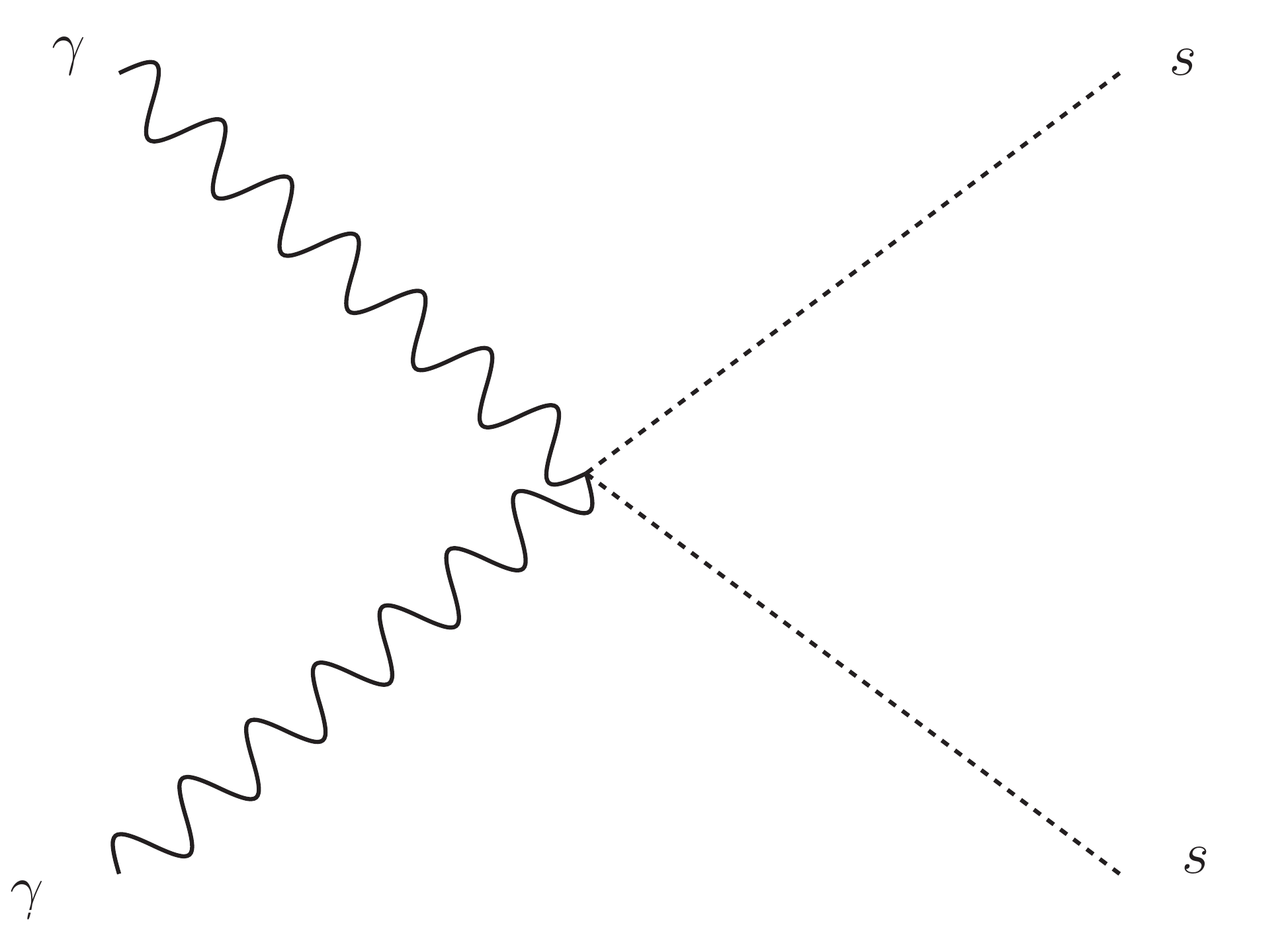}
\includegraphics[width=.4\textwidth]{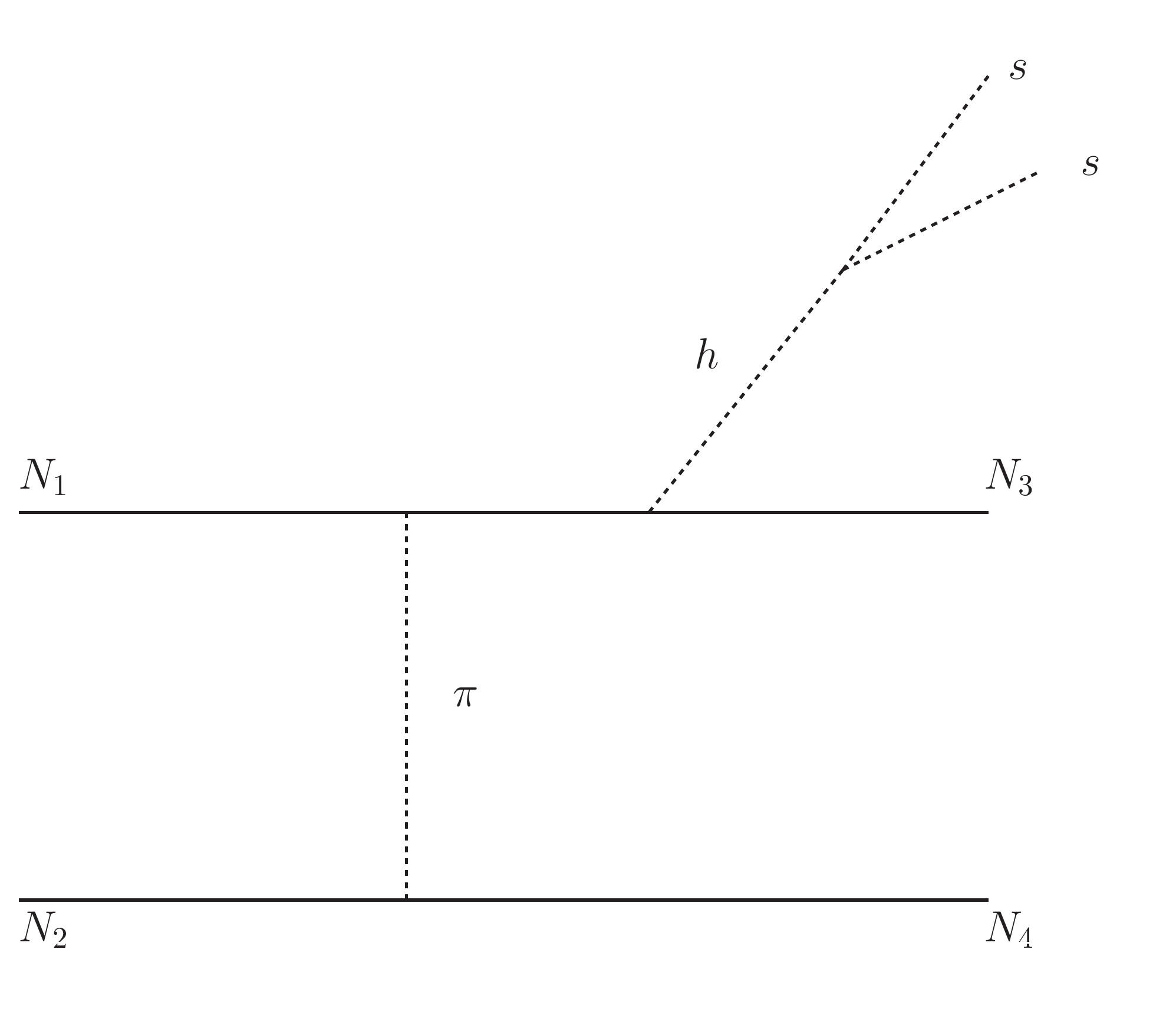}
\caption{\label{fig:SNemission}
 Photon annihilation (left) and bremsstrahlung-like scalar emission in
nucleon-nucleon interactions. For the simplified model, the Higgs has
to be replaced by the new mediator $\phi$ in the right plot.}
\end{figure}

\paragraph{Scalar dark matter} 
Following Ref.~\cite{Olive:2007aj}, we consider two processes for the
energy-loss rate in the free-streaming limit, photon
annihilation $\gamma\gamma \to s s$ and the bremsstrahlung-like
process $N N \to N N s s$ both depicted in Fig.~\ref{fig:SNemission}.
Pair annihilations of photons yields a cross section of
\begin{align}
\sigma_{\gamma \gamma \to s s}
\approx
\frac{ \lambda_{hs}^2g_{h\gamma\gamma}^2}{\pi}  \, \frac{\omega^2}{m_h^4} \; ,
\label{eq:ggann_xsec}
\end{align}
where $\omega$ is the energy of the incoming photon.  For the total
energy loss for a thermalized gas of photons, we
obtain~\cite{Olive:2007aj}
\begin{align}
 \Gamma_{\gamma\gamma}=
  n_{\gamma}^2\langle 2\omega
\sigma_{\gamma\gamma\to ss}\rangle =
\frac{32\pi\zeta(3)}{63}\lambda_{hs}^2g_{h\gamma\gamma}^2\frac{T_\text{SN}^9}{m_h^4}\,,
\end{align}
where
\begin{align}
n_\gamma&=\frac{2\zeta(3)}{\pi^2}T_\text{SN}^3\notag \,, \\[1.5mm]
\langle 2\omega \; \sigma_{\gamma\gamma\to ss}\rangle
        &=\frac{\int_0^\infty d\omega \omega^2 \; 2 \omega
          \sigma_{\gamma\gamma\to ss}/[\exp(\omega/T_\text{SN})-1]}{\int_0^\infty
          d \omega \omega^2 \; 1/[\exp(\omega/T_\text{SN})-1]} \; .
\end{align}
With  the temperature given
in Eq.\eqref{eq:coreval} we can translate Eq.\eqref{eq:SNlimit} into 
\begin{align}
\lambda_{h s} < 0.17 \; .
\end{align}
The second process $N N \to N N s s$ is calculated with an effective
nucleon-scalar coupling. We can simply translate limits on $M_*$ in
Ref.~\cite{Olive:2007aj},
\begin{align}
\langle E\sigma v\rangle &\sim
\frac{1}{12\pi^4}\frac{T^3m_N^2}{M_{*}^4}\left(\frac{T}{m_N}\right)^{1/2}\sigma_{NN}\notag \\
\Rightarrow \qquad \Gamma_{NN \to NNss} &\sim
  \sigma_{NN \to NNss}\frac{n_N^2T^{7/2}m_N^{3/2}}{12\pi^4M_{*}^4} < 10^{-14}~\text{MeV}^5\,,
\end{align}
through $c_{sNN} = m_N/(2\,M_*^2)$ into the model parameters for scalar
ULDM
\begin{alignat}{7}
\lambda_{h s} &< 2.75 \cdot 10^{-4}
&\quad &\text{(Higgs portal)}\,, \notag \\
\frac{\mu_{\phi s}}{\Lambda_\phi} &< 1.8\cdot 10^{-5}\left(\frac{M_S}{100\, \text{GeV}}\right)^2
&\quad &\text{(scalar mediator)} \; .
\end{alignat}
For the Higgs portal we see that the constraints from $NN\to NNss$
scattering are stronger than those from photon annihilation. In the
following, we therefore omit the annihilation process.

For the trapping limit we assume the nuclei in the supernova to be
at rest, because $T_{SN}\sim30$ MeV $\ll M_N$. The
elastic scattering cross section is given by
\begin{align}
\sigma_{s N \to s N} & = \frac{1}{8\pi} \ 
c_{sNN}^2 \;
\frac{2 +4 \frac{E_s}{M_N} + \frac{E_s^2}{M_N^2} + \frac{m_s^2}{M_N^2} }{\left(  1 + 2 \frac{E_s}{M_N} + \frac{m_s^2}{M_N^2}  \right)^2}
\simeq \frac{1}{4\pi} \; 
c_{sNN}^2\,,
\end{align}
where we assume $m_\phi\ll M_N$ as well as $E_\phi\sim T_{SN} \ll
M_N$. The coupling strength $c_{sNN}$ for the scalar models can be
found in Eq.\eqref{eq:ScalarNN}. With that cross section, we find a
constraint on the coupling and suppression scale through the optical
depth criterion
\begin{alignat}{7}
\lambda_{h s}&> 6.7~...~8.1\cdot 10^{-3}
&\quad &\text{(Higgs portal)} \,,\notag \\
\frac{\mu_{\phi s}}{\Lambda_\phi} &> 4.3~...~5.2\cdot 10^{-4}\left(\frac{m_\phi}{100~\text{GeV}}\right)^2
&\quad &\text{(scalar mediator)} \; .
\end{alignat}
For all limits in the trapping regime, we state a range for the
constraint that is coming from the variable choice of $m$ in the temperature and
density profile. In our final plots, we choose the trapping limit that
maximizes the excluded bands for supernova constraints.

\paragraph{Pseudoscalar dark matter}
For the derivative portal models, we start with the energy-loss rate
from Ref.~\cite{Keung:2013mfa}
\begin{align}
\Gamma_{NN\to NNaa}=\varepsilon_x\rho_\text{core}\simeq \frac{264\sqrt{\pi}}{\pi^4}\left(3-\frac{2\beta}{3}\right)n_B^2\left(\frac{g_Nm_N}{m_r^2m_h^2}\right)^2\alpha_\pi^2\frac{T^{9.5}}{m_N^{4.5}} \; ,
\end{align}
where $\alpha_\pi=(2m_Nf/m_\pi)^2/4\pi$ with $f\approx 1$ is the
pion-nucleon ``fine-structure'' constant, $g_Nm_N/(m_r^2m_h^2)$ is the
effective Higgs-nucleon coupling and $\beta$ is a nucleon-momentum
dependent term with $\beta=2.0938$~\cite{Keung:2013mfa}. We replace
the effective $\alpha N$-coupling with the effective $aN$-coupling
$g_Nm_N/(m_r^2m_h^2) \to c_{aNN}$, and obtain the limits on our model
parameters
\begin{alignat}{7}
\frac{1}{\Lambda_{ha}} &< \frac{0.3}{\text{GeV}} 
&\quad &\text{(Higgs mediator)}\,, \notag \\
\frac{1}{\Lambda_\phi} &< \frac{6.3\cdot
                         10^{-2}}{\text{GeV}} \left(\frac{\Lambda_{\phi
    a}}{10~\text{GeV}}\right) \left(\frac{m_\phi}{100~\text{GeV}}\right)
&\quad &\text{(scalar mediator)} \; .
\end{alignat}

For the squared matrix element in the derivative simplified model, we
get
\begin{align}
|\mathcal{M}|^2=\frac{1}{8}c_{aNN}^2(2m_a^2-t)^2(4m_N^2-t).
\end{align}
By assuming $m_a\ll E_a \ll m_N$, this yields a cross section of
\begin{align}
\sigma_{aN\to aN}=\frac{c_{aNN}^2}{12\pi}T^4\,,
\end{align}
where $c_{aNN}$ is taken from Eq.~\ref{eq:DerNN}.
Following the calculations of the scalar case, the limits in the
derivative portal models are
\begin{align}
\frac{1}{\Lambda_{ha}} &>17.1-34.1~\text{GeV}^{-1}\notag \,,\\
\frac{1}{\Lambda_\phi} &> 2.0-7.4\cdot
                         10^2~\text{GeV}^{-1}\left(\frac{\Lambda_{\phi
                         a}}{10~\text{GeV}}\right) \left(\frac{m_\phi}{100~\text{GeV}}\right)^2~.
\end{align}
%

\subsection{Direct detection}

In all our DM scenarios the scalar DM candidate couples to nucleons
via a scalar mediator, either the SM Higgs $h$ or a new singlet
$\phi$. At low energies these interactions will lead to elastic
DM-nucleus scattering, where the incoming DM particle will transfer
part of its momentum to the nucleus and generate a nuclear recoil. The
corresponding spectrum depends on the DM-nucleon scattering cross
section~\cite{Cerdeno:2010jj,Gelmini:2015zpa,Lin:2019uvt}.  In all
cases where the DM-nucleon interactions carry a scalar Lorentz
structure, the induced scattering is spin-independent.

\paragraph{Scalar Higgs portal}
In a first step we derive the relevant effective low-energy
operator. For the scalar Higgs portal it is generated when we
integrate out the Higgs in Eq.\eqref{eq:higgsportal} and take into
account the Higgs-quark coupling $m_q/v\, h \, \bar q q$,
\begin{align}
    \lag \supset \frac{\lambda_{hs}\, m_q}{2\, m_h^2} \, s^2\, \bar q q\,.
\end{align}
To obtain the scattering cross section of the DM particle with
nucleons we evaluate the matrix element of the effective (partonic)
operator with asymptotic nucleon states,
\begin{align}
    \mathcal{M} = \langle n(k')| \mathcal{O}_\text{eff} | n(k) \rangle  \,.
\end{align}
For light quarks we evaluate the nuclear matrix
element~\cite{Plehn:2017fdg},
\begin{align}\label{eq:mnuc_light}
      \langle n(k')| m_q\, \bar q q  | n(k) \rangle   = m_n\ f^n_{T,q} \ \bar u _n(k') \, u_n(k)\; ,
\end{align}
while for heavy quarks we will integrate out the heavy quark fields
via the QCD trace anomaly~\cite{Lin:2019uvt}, leading for each heavy
quark field to the replacement
\begin{align}
        m_q \, \bar q q \rightarrow - \frac{\alpha_s}{12 \pi} \Tr G_{\mu\nu}G^{\mu\nu} \,.
\end{align}
The nuclear matrix element can then be evaluated via
\begin{align}
        \langle n(k')| \alpha_s\,  \Tr G_{\mu\nu}G^{\mu\nu} | n(k) \rangle   = - \frac{8 \pi}{9} \, m_n\ f^n_{T,g}\  \bar u _n(k') \, u_n(k)\,. \label{eq:nucl_gluon}
\end{align}
The coefficients $f^n_{T,q}$ and $f^n_{T,g}$ parametrize the nuclear
matrix elements and can be found \eg in Table II of
Ref.~\cite{Lin:2019uvt}. Summing over all the quarks, the full matrix
element for the DM-nucleon scattering reads
\begin{align}
    \mathcal{M} =& \frac{\lambda_{hs}}{m_h^2} \, m_n\, \left( f^n_{T,u} + f^n_{T,d} + f^n_{T,s} + \frac{2}{9}\, f^n_{T,g}\right) \ \bar u _n(k') \, u_n(k)\,.
\end{align}
To turn this into a cross section section for DM-nucleon scattering we
perform a non-relativistic expansion of the matrix element. This is a
good approximation as long as the scattered nuclei are at rest and the
DM particles in the local halo are non-relativistic. In the
non-relativistic limit the we can take the normalization of the Dirac
spinors as $u(k)^s=\sqrt{m}\ (\xi_s,\xi_s)^T$, where $\xi_s$ are the
two-component Weyl spinors normalized such that $\xi_s^\dagger
\xi_s=1$.  This allows us to write the non-relativistic nuclear matrix
element as
\begin{align}
    \mat_{NR} =& \frac{\lambda_{hs}\, m_n}{m_h^2} \, \left( f^n_{T,u} + f^n_{T,d} + f^n_{T,s} + \frac{2}{9}\, f^n_{T,g}\right) \ (2m_n)\,  \xi^\dagger_{s'} \xi_s\,.
\end{align}
Summing and averaging over  the nucleon spins we arrive at the squared matrix element,
\begin{align}
  |\mat |^2 = \left(\frac{2\,\lambda_{hs}\, m_n^2}{m_h^2}\right)^2 \, \left( f^n_{T,u} + f^n_{T,d} + f^n_{T,s} + \frac{2}{9}\, f^n_{T,g}\right)^2\,.
\label{eq:MDDHP}
\end{align}
If the squared matrix element is independent of the scattering angle
we can trivially integrate over the scattering angle to obtain the
total scattering cross section
\begin{align}
    \sigma_{sn \to sn} = \frac{1}{\pi} \,\left( \frac{\mu_{sn}\,\lambda_{hs}\,m_n}{2\,m_s\,m_h^2}\right)^2 \ \left( f^n_{T,u} + f^n_{T,d} + f^n_{T,s} + \frac{2}{9}\, f^n_{T,g} \right)^2\,,
\end{align}
with the reduced DM-nucleon mass $\mu_{sn}=m_s\,m_n/(m_s+m_n)$.

\paragraph{Scalar with new mediator}
The case of a scalar DM particle $s$ coupling to a new scalar mediator
$\phi$ described by Eq~\eqref{eq:scalarlag} differs mainly by the
interactions of $\phi$ with matter. In contrast to the Higgs portal
model, the interaction is mediated via the effective coupling to
gluons rather than the renormalizable couplings to quarks.  We find
the relevant effective operator at energy scales much below the
mediator mass $m_\phi$ to be
\begin{align}
    \lag \supset \frac{\alpha_s\, \mu_{\phi s}}{2 \Lambda_\phi\, m_{\phi}^2}\ s^2\ \Tr G_{\mu\nu} G^{\mu\nu} \,.
\end{align}
Using the result for the nuclear matrix element of the gluon operator
from Eq.\eqref{eq:nucl_gluon} we can write down the
nuclear-level matrix element for the DM-gluon contact term,
\begin{align}
    \mat = - \frac{8 \pi}{9} \frac{m_n\,  \mu_{\phi s}}{\Lambda_\phi\, m_\phi^2} \ f^n_{T,g} \ \bar u _n(k') \, u_n(k)\,.
\end{align}
Repeating the same steps as for the Higgs
portal model the DM-nucleon elastic scattering cross section becomes
\begin{align}
    \sigma_{sn \to sn} = \frac{16 \pi }{81}\ \left(\frac{\mu_{sn}\,m_n \, \mu_{\phi s}}{\Lambda_\phi\, m_s\,m_\phi^2}\right)^2 \ (f^n_{T,g})^2\,.
\end{align}

\paragraph{Pseudoscalar with Higgs mediator}
When the pseudoscalar $a$ couples to the Higgs via the derivative
portal of Eq.\eqref{eq:derivativePortal}, integrating out the Higgs
yields a momentum-dependent contact term,
\begin{align}
  \lag \supset \frac{m_q}{m_h^2 \Lambda_{ha}^2} \frac{p\cdot p'}{2}  \; a^2 \bar{q}q \; .
\end{align}
The corresponding nuclear matrix element for DM-nucleon scattering
with asymptotic nucleon initial and final states is
\begin{align}
    \mat = \frac{p\cdot p'}{m_h^2 \, \Lambda_{ha}^2}\,m_n\ \left( f^n_{T,u} + f^n_{T,d} + f^n_{T,s} + \frac{2}{9}\, f^n_{T,g}\right) \ \bar u _n(k') \, u_n(k)\,.
\end{align}
In terms of the reduced mass $\mu_{an}=m_a\,m_n/(m_a+m_n)$ we arrive
at the full expression for total cross section in the non-relativistic
limit,
\begin{align}
    \sigma_{na \to na} = \frac{\mu_{an}^2\, m_a^2\,m_n^2}{4\pi\,\Lambda_{ah}^4\,m_h^4} \left[1- 2\left(1-\frac{\mu_{an}}{m_a}\right)v_a^2+\left(1-2\frac{\mu_{an}}{m_a}+\frac{4}{3}\frac{\mu_{an}^2}{m_a^2}\right)v_a^4\right] \notag \\
    \times\left( f^n_{T,u} + f^n_{T,d} + f^n_{T,s} + \frac{2}{9}\, f^n_{T,g}\right)^2\; ,
\end{align}
where  the higher power terms in $v_a\sim10^{-3}$ can be ignored.

\paragraph{Pseudoscalar with new mediator}
Lastly, in the case of a derivative portal coupling to a new scalar
mediator, Eq.\eqref{eq:scalarALPs}, we obtain the low-energy contact
term
\begin{align}
    \lag \supset - \frac{\alpha_s}{\Lambda_{\phi a} \Lambda_\phi  m_\phi^2} \frac{p\cdot p'}{2} \; a^2 \Tr G_{\mu\nu}G^{\mu\nu} \,.
\end{align}
Evaluating the matrix element with asymptotic nucleon initial and
final states as before and performing a non-relativistic expansion we
obtain the scattering rate
\begin{align}
    \sigma_{na \to na} = \frac{16\pi}{81}\frac{\mu_{an}^2\, m_a^2\,m_n^2}{\Lambda_{ah}^2\,\Lambda_\phi^2\,m_h^4} \left[1- 2\left(1-\frac{\mu_{an}}{m_a}\right)v_a^2+\left(1-2\frac{\mu_{an}}{m_a}+\frac{4}{3}\frac{\mu_{an}^2}{m_a^2}\right)v_a^4\right] \left(  f^n_{T,g}\right)^2\,.
\end{align}

\subsection{FCC projections}
\label{app:fcc}

\begin{figure}[t]
\includegraphics[width=.5\textwidth]{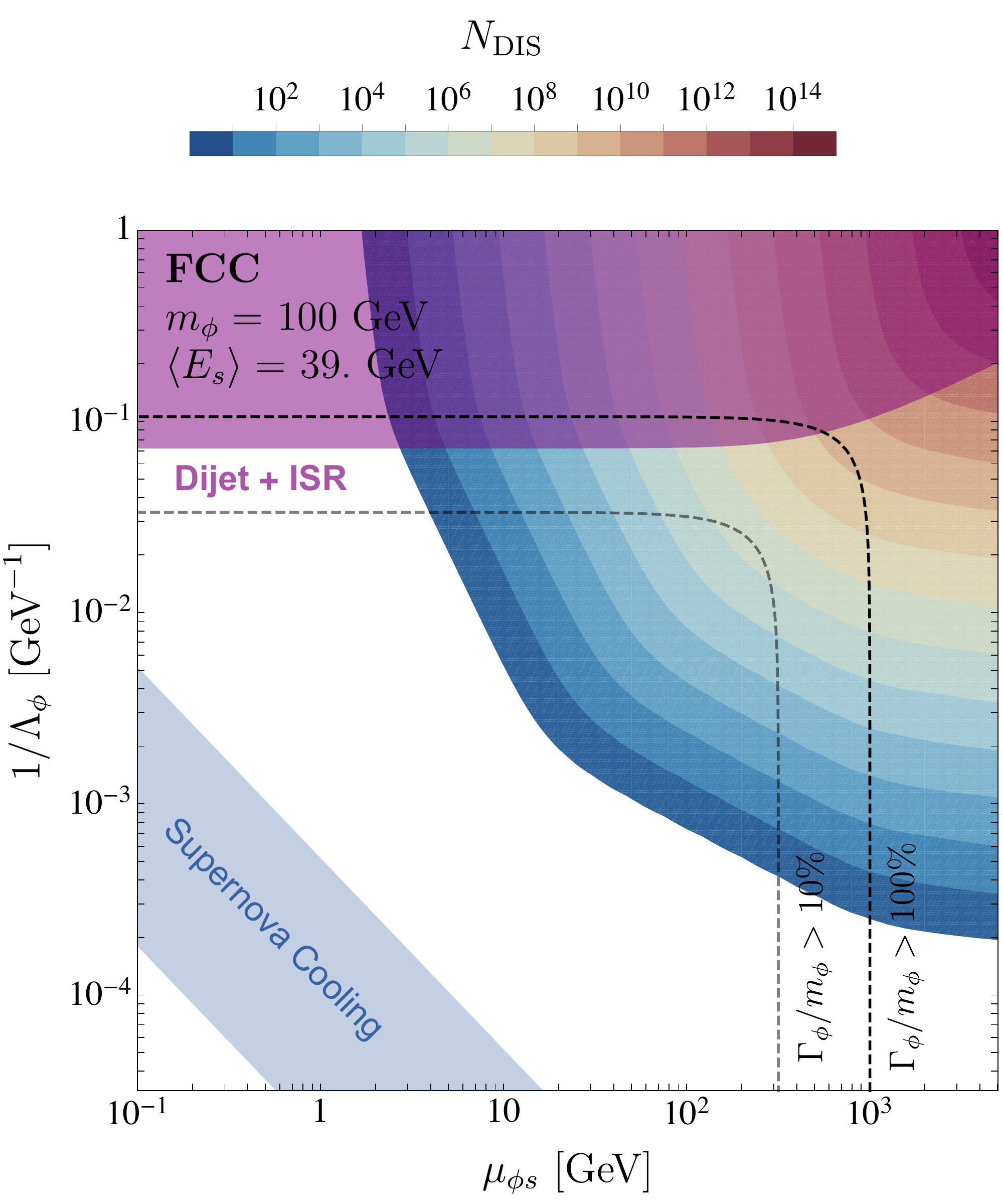}%
\includegraphics[width=.5\textwidth]{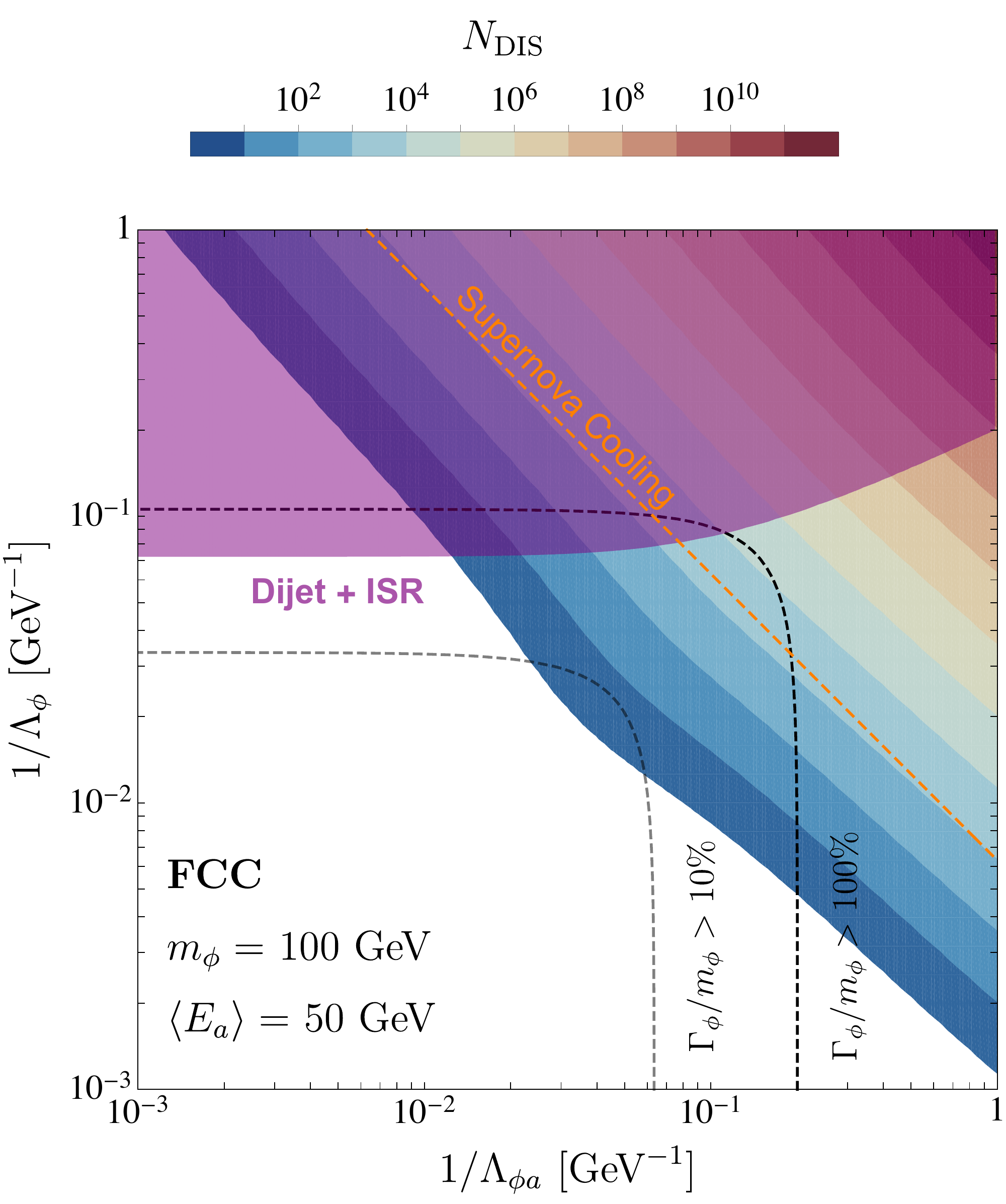}
\caption{Left: number of expected DIS events at the FCChh in the plane
  of DM-mediator coupling $\mu_{\phi s}$ versus mediator-gluon
  coupling $\Lambda_\phi$ for scalar ULDM with $m_\phi
  =100$~GeV. Right: same for the pseudoscalar ULDM.}
\label{fig:NdisFCC} 
\end{figure}

Complementing our LHC analysis of DIS in scalar simplified DM models
at the LHC in Sec.~\ref{sec:LHC_DD}, we illustrate our results for a
future circular hadron collider (FCChh) with an energy of 100~TeV and
an integrated luminosity of $30~\iab$~\cite{Benedikt:2018csr}.  As for
the LHC we generate Monte Carlo events using
\texttt{MadGraph5\_aMC@NLO}~\cite{Alwall:2014hca} for the DM
production channels 
\begin{align}
  p p \to \phi \to s s \; , \; a a
\end{align}
The mediator $\phi$ couples to gluons through the usual dimension-5
operator. For these signals we calculate the number of expected DIS
events as outlined in Sec.~\ref{sec:LHC_DD}. We illustrate our result
in Fig.~\ref{fig:NdisFCC}. Comparing our findings for the FCC in the
simplified portal model in the left panel of Fig.~\ref{fig:NdisFCC}
with those for the LHC shown in the left panel of
Fig.~\ref{fig:NdisSSimp}, we see that at the FCC we expect roughly a
factor of hundred more events for a given point in the parameter
space.  For the pseudoscalar, derivative case we compare the right
panel of Fig.~\ref{fig:NdisFCC} to the left panel of
Fig.~\ref{fig:NdisDSimp} and find a thousand times as many events for
the FCC.  This is expected from the momentum enhancement shown in
Eq.\eqref{eq:disSSimp} compared to Eq.\eqref{eq:disDSimp}.

\end{fmffile}
\bibliography{literature}

\end{document}